\documentclass[pre,twocolumn,showpacs,floatfix]{revtex4}
\usepackage{graphicx}
\usepackage{amsmath}
\usepackage{amssymb}
\usepackage{subfigure}
\newcommand{\ww}[1]{\underline{\underline{{\bf #1}}}}

\newcommand{\be}{\begin{equation}}
\newcommand{\ee}{\end{equation}}
\newcommand{\bea}{\begin{eqnarray}}
\newcommand{\eea}{\end{eqnarray}}
\newcommand{\ba}{\begin{aligned}}
\newcommand{\ea}{\end{aligned}}
\newcommand{\bma}{\begin{bmatrix}}
\newcommand{\ema}{\end{bmatrix}}
\newcommand{\Pp}{{\mathcal P}}
\newcommand{\kk}{{\ww{\mathcal K}}}
\newcommand{\GG}{{\ww{\mathcal G}}}

\newcommand{\fext}{{\bf F}^{\text{ext}}}
\newcommand{\sti }{\ww{K}}
\newcommand{\stia }{\sti^{(1)}}
\newcommand{\stib }{\sti^{(2)}}
\newcommand{\trs}[1]{\hspace{.05em}\,^{\bf T}\hspace{-.1em}\ww{#1}}
\newcommand{\rig }{\ww{G}}
\newcommand{\rigt }{\trs{G}}
\newcommand{\norm}[1]{\vert \vert {\bf #1}\vert \vert}
\newcommand{\normm}[1]{\vert \vert #1\vert \vert}
\newcommand{\bi}{\begin{itemize}}
\newcommand{\ei}{\end{itemize}}
\newcommand{\im}{\item}
\newcommand{\nij}{{\bf n}_{ij}}
\newcommand{\nik}{{\bf n}_{ik}}
\newcommand{\Tij}{{\bf T}_{ij}}
\newcommand{\hhh}{\textsc{h}}

\newcommand{\pk}{\ww{\pi}}

\newcommand{\tr}{\text{tr}\hspace{.05em}}
\newcommand{\disty}{\displaystyle}
\newcommand{\ave}[1]{\langle #1 \rangle}
\begin{document}
\title{Internal states of model isotropic granular packings. \\
I. Assembling process, geometry and contact networks.}

\author{Ivana Agnolin\footnote{Present address: Geoforschungszentrum, 
Haus D, Telegrafenberg, D-14473 Potsdam, Germany}
}
\author{Jean-No\"el Roux}
\email{jean-noel.roux@lcpc.fr}
\affiliation{Laboratoire des Mat\'eriaux et des Structures du G\'enie Civil\footnote{
LMSGC is a joint laboratory depending on Laboratoire Central des Ponts et Chauss\'ees, \'Ecole Nationale
des Ponts et Chauss\'ees and Centre National de la Recherche Scientifique},
Institut Navier, 2 all\'ee Kepler, Cit\'e Descartes, 77420 Champs-sur-Marne, France}

\date{\today}

\begin{abstract}
This is the first paper of a series of three, in which we
report on numerical simulation studies of 
geometric and mechanical properties of static assemblies of spherical beads under an isotropic pressure. 
The influence of various assembling processes on packing microstructures is investigated. It is accurately checked that
frictionless systems assemble in the unique random close packing (RCP) state in the low pressure limit 
if the compression process is fast  enough, higher solid fractions corresponding to more ordered 
configurations with traces of crystallization. Specific properties directly related to isostaticity of
the force-carrying structure in the rigid limit are discussed. With frictional grains,
different preparation procedures result in quite different inner structures that cannot be classified by the sole density. 
If partly or completely lubricated they will assemble like frictionless ones, approaching the RCP
solid fraction $\Phi_{\text{RCP}}\simeq 0.639$ 
with a high coordination number: $z^*\simeq 6$ on the force-carrying backbone. If compressed with a realistic
coefficient of friction $\mu=0.3$ packings stabilize in a loose state with $\Phi\simeq 0.593$ and $z^*\simeq 4.5$. And, more surprisingly,  
an idealized ``vibration'' procedure, which maintains an agitated, collisional r\'egime up
to high densities results in equally small values of $z^*$ while $\Phi$ is close to the maximum value $\Phi_{\text{RCP}}$. 
Low coordination packings have a large proportion ($>$10\%) of rattlers -- grains carrying no force -- the effect of
which should be accounted for on studying position correlations, and also contain a small proportion of 
localized ``floppy modes'' associated with divalent grains. 
Low pressure states of frictional packings retain a finite level of force indeterminacy even when assembled with the slowest compression rates
simulated, except in the case when the friction coefficient tends to infinity.
Different microstructures are characterized in terms of near neighbor correlations on various scales, 
and some comparisons with available laboratory data are reported, although values of contact 
coordination numbers apparently remain experimentally inaccessible. 
\end{abstract}


\maketitle

\section{Introduction}
\subsection{Context and motivations}
The mechanical properties of solidlike granular packings and their microscopic, grain-level origins
are an active field of research in material science and condensed-matter physics~\cite{HHL98,HW04,GRMH05}.
Motivations are practical, originated in  soil mechanics and material processing, as well as
theoretical, as general approaches to the rheology of different physical systems made of particle assemblies out of
thermal equilibrium~\cite{LN01} are attempted.

The packing of equal-sized spherical balls is a simple model for which 
there is a long tradition of \emph{geometric} characterization studies. Packings are usually classified according to
their density or solid volume fraction $\Phi$, and the frequency of occurrence of some local patterns. 
Direct observations of packing microstructure is difficult, although it has recently benefitted from
powerful imaging techniques~\cite{RPBBTB03,ASSS04,AsSaSe05}. 
The concept of random close packing (RCP), is often invoked~\cite{CC87,BD91}, although some
authors criticized it as ill-defined~\cite{TTD00}. 
It corresponds to the common observation that bead packings without any trace of
crystalline order do not exceed a maximum density, $\Phi_{RCP}$, slightly below $0.64$~\cite{CC87}. 

\emph{Mechanical} studies in the laboratory have been performed on granular 
materials for decades in the realm of soil mechanics, 
and the importance of packing fraction $\Phi$ on the rheological 
behavior has long been recognized~\cite{DMWood,BiHi93,MIT93,HHL98}. 
The anisotropy of the packing microstructure, due to the assembling process,
has also been investigated~\cite{Oda72,AM72}, and shown to influence the stress-strain behavior of test samples~\cite{Tat33}, as well
as the stress field and the response to perturbations of gravity-stabilized sandpiles or granular layers~\cite{VHCBC99}.

Discrete numerical simulation~\cite{CS79} proved a valuable tool to investigate the internal state of packings, 
as it is able to reproduce mechanical behaviors, 
and to identify relevant variables other than $\Phi$, such as coordination number and fabric
(or distribution of contact orientations)~\cite{BR90,RR01,RR04,RK04,RoCh05}. 
In the case of sphere
packings, simulations have been used to characterize the geometry of gravity-deposited
systems~\cite{SEGHL02,SGL02} or oedometrically compressed ones~\cite{MJS00}, 
to investigate the quasistatic, hysteretic stress-strain dependence in solid packings~\cite{TH00,SUFL04}, 
and their pressure-dependent elastic moduli in a compression experiment~\cite{MGJS99,Makse04}. 

However, in spite of recent progress, quite a few basic questions remain unresolved.
It is not obvious how closely the samples used in numerical
simulations actually resemble laboratory ones, for which density is often the only available state parameter.
Both simulations and experiments resort
to certain preparation procedures to assemble granular packings, which,
although their influence is recognized as important, are seldom
studied, or even specified. One method to produce dense packings in simulations is to set
the coefficient of intergranular friction to zero
~\cite{TH00,MJS00} or to a low value~\cite{SUFL04} in the assembling stage,
while a granular gas gets compacted and equilibrated
under pressure. On keeping frictionless contacts, 
this results, in the limit of low confining stress,
in dense systems with rather specific properties~\cite{OSLN03,Makse04},
related to isostaticity and potential energy minimization~\cite{JNR2000}. 
Examples of traditional procedures in soil mechanics are rain deposition under gravity, also known as air pluviation 
(which produces homogeneous states if grain flow rate and height of
free fall~\cite{RT87} are maintained constant) and layerwise deposition and dry or moist tamping. Those two methods were observed to produce,
in the case of loose sands,
 different structures for the same packing density~\cite{BeCaDu04}. 
Densely packed particle assemblies can also be obtained in the laboratory by vibration, or application of 
repeated ``taps''~\cite{NKBJN98,PB02} to a loose deposit.
How close are dense experimental sphere packings to model configurations obtained on simulating 
frictionless particles ? How do micromechanical parameters influence the packing structure ?
Is the low pressure limit singular in laboratory grain packings and in what sense ?
\subsection{Outline of the present study}
The present paper provides some answers to such questions, from numerical simulations in
the simple case of isotropically assembled and
compressed homogeneous packings of spherical particles. It is the first 
one of a series of three, and deals with the geometric characterization of low pressure
isotropic states assembled by different procedures, both without and with intergranular friction.
The other two, hereafter referred to as papers II~\cite{iviso2} and III~\cite{iviso3}, respectively investigate the effects of compressions and pressure cycles, and
the elastic response of the different numerical packings, with comparisons to experimental results. 
Although mechanical aspects are hardly dealt with in the present paper, we insist that geometry and mechanics are strongly and mutually related.
We  focus here on the variability of the coordination numbers, which will prove important for mechanical response properties of granular packings, and
show that equilibrated packs of identical beads can have a relatively large numbers of ``rattler'' grains, which do not participate in
force transmission.  We
investigate the dependence of initial states on the
assembling procedure, both with and without friction.
We study the effects of procedures designed to produce dense states
(close to RCP), and we characterize the geometry of such states on different scales. 

It should be emphasized that we do not claim here to mimic experimental
assembling procedures very closely. Rather, we investigate the results of several preparation methods, which are
computationnally convenient, maintain isotropy, and produce equilibrated samples with rather different characteristics. Those
methods nevertheless share some important features with laboratory procedures, and we shall argue that the resulting
states are plausible models for experimental samples. 

The numerical model and the simulation procedures (geometric and mechanical parameters, contact law,
boundary conditions) are presented in Section~\ref{sec:model}, where some basic definitions and
mechanical properties pertaining to granular packs are also
presented or recalled. Part~\ref{sec:asf} discusses the properties of 
frictionless packings, and introduces several characterization approaches used in the general case as well.
Section~\ref{sec:assemblef} then describes different assembling procedures of frictional packings and the resulting 
microstructures. Section~\ref{sec:concl} discusses  perspectives to the present study, some of which are pursued in papers II and III of the series.
Appendices deal with technical issues, and also present a more detailed comparison 
with some experimental data.

This being a long paper, it might be helpful to specify which parts can be read independently.
On first going through the paper, the reader might skip Section~\ref{sec:cryst}, dealing with a rather specific issue.
The properties stated or recalled in Section~\ref{sec:rist} are used to discuss stability issues and 
isostatic values of coordination numbers, but they can also be overlooked in a first approach. Finally, Section~\ref{sec:assemblef} can be read independently from
Section~\ref{sec:asf}, apart from the explanations about equilibrium conditions (in paragraph~\ref{sec:MDres}) and the treatment of rattlers (paragraph~\ref{sec:ratt}). 
Sections~\ref{sec:asf} and~\ref{sec:assemblef} both have conclusive subsections which summarize the essential results. 
\section{Model, numerical procedures, basic definitions\label{sec:model}}
\subsection{Intergranular forces \label{sec:forces}}
We consider spherical beads of diameter $a$ (the value of which, as we ignore gravity, will prove irrelevant),
interacting in their contacts by
the Hertz law, relating the normal force $N$ to the elastic normal deflection $h$ as~:
\be
N=\frac{\tilde E\sqrt{a}}{3}h^{3/2}.
\label{eqn:hertz}
\ee
In Eqn.~\ref{eqn:hertz}, we introduced the notation 
$$
\tilde E = \frac{E}{1-\nu^2}, 
$$
$E$ is the Young modulus of the beads, and $\nu$ the Poisson ratio. 
For spheres, $h$, the elastic deflection of the contact, is simply the distance of approach of the centers beyond the first contact.
The normal stiffness $K_N$ of the contact is defined as the rate of change of the force with normal displacement:
\be
K_N = \frac{dN}{dh} = \frac{\tilde E\sqrt{a}}{2}h^{1/2} = \frac{3^{1/3}}{2} \tilde E^{2/3} a^{1/3} N^{1/3}
\label{eqn:kn}
\ee
Although many geometric features of particle packings do not depend on the details of the model for contact elasticity, 
and could be observed as well with a simpler, linear unilateral elastic model, it is necessary to implement suitable non-linear
contact models to deal with the mechanical properties in papers II and III~\cite{iviso2,iviso3}. Tangential elasticity and friction
in contacts are appropriately described by the Cattaneo-Mindlin-Deresiewicz laws~\cite{JO85}, which we implement in a
simplified form, as used \emph{e.g.}, in refs.~\cite{MGJS99,SRSvHvS05}: 
the tangential stiffness $K_T$ relating, in the elastic regime,
the increment of tangential reaction $d{\bf T}$ to the 
relative tangential displacement increment $d{\bf u}_T$ is a function of $h$ (or $N$) alone
(\emph{i.e.,} it is kept constant, equal to its 
value for ${\bf T}=0$): 
\be
\ba
d{\bf T} &= K_T d{\bf u}_T,\ \ \mbox{with}\\
K_T&=\frac{2-2\nu}{2-\nu} K_N=\frac{1-\nu}{2-\nu}\tilde E\sqrt{a}h^{1/2}
\end{aligned}
\label{eqn:tang}
\ee
To enforce the Coulomb condition with friction coefficient $\mu$,
${\bf T}$ has to be projected back onto the circle of radius $\mu N$
in the tangential plane whenever the increment
given by eqn.~\ref{eqn:tang} would cause its magnitude to exceed this limit. Moreover, when 
$N$ decreases to $N-\delta N$, ${\bf T}$ is scaled down to the value
it would have had if $N$ had constantly been equal to $N-\delta N$ in the past.
It is not scaled up when $N$ increases.
Such a procedure, suggested \emph{e.g.}, in~\cite{EB96}, avoids spurious increases of
elastic energy for certain loading histories. More details are given in
Appendix~\ref{sec:appendixfnft}.

Finally, tangential contact forces have to follow the material motion.
Their magnitudes are assumed here not to be affected by rolling 
(\emph{i.e.}, rotation about a tangential axis) or
pivoting (\emph{i.e.}, rotation about the normal axis), while their direction rotates
with the normal vector due to rolling,
and spins around it with the average spinning rate of the two spheres (to ensure objectivity).
The corresponding equations are 
given in Appendix~\ref{sec:appendixTrot}.

In addition to the contact forces specified above, we introduce viscous ones, which oppose the normal
relative displacements (we use the convention that positive normal forces are repulsive):
\be
N^v = \alpha(h) \dot h
\label{eqn:fvisc}
\ee 
The damping coefficient $\alpha$ depends on $h$, and
we choose its value as a fixed fraction $\zeta$ of the critical damping coefficient of the normal (linear) 
spring of stiffness $K_N(h)$ (as given by~\eqref{eqn:kn}) joining two beads of mass $m$:
\be
\alpha(h) = \zeta \sqrt{2mK_N(h)}.
\label{eqn:defzeta}
\ee
From~\eqref{eqn:kn}, $\alpha$ is thus proportional to $h^{1/4}$, or to $N^{1/6}$. The same damping law was used in~\cite{SRSvHvS05}.
Admittedly, the dissipation given by \eqref{eqn:fvisc}-\eqref{eqn:defzeta} has little physical justification,
and is rather motivated by computational convenience. We shall therefore assess the influence of $\zeta$ on the numerical results. 
The present study being focussed on statics, we generally use a strong dissipation, $\zeta = 0.98$, to approach equilibrium faster.
This particular value is admittedly rather arbitrary: the initial motivation for choosing $\zeta<1$ is the computational inefficiency 
of overdamped contacts with $\zeta >1$ in the case of \emph{linear contact elasticity}. Yet we did not check whether values of 1 or 
even higher would cause any problem \emph{with Hertzian contacts}. 
In the linear case, the restitution coefficient in a binary collision varies as a very
fastly decreasing function of $\zeta$, and changes of $\zeta$ in the range between $0.7$ and 1 have virtually no detectable effect.

We do not introduce any tangential viscous force, and impose the Coulomb inequality to elastic force components
only.  
We choose the elastic parameters $E=70$~GPa and $\nu=0.3$, suitable for glass beads, and the friction coefficient
is attributed a moderate, plausible value $\mu=0.3$. 
These choices are motivated by comparisons to experimental measurements of elastic moduli, 
to be carried out in paper III~\cite{iviso3}.
 \subsection{Boundary conditions and stress control\label{sec:bcsc}}
The numerical results presented below were obtained on samples of $n=4000$ beads,
enclosed in a cubic or parallelipipedic cell with
periodic boundary conditions.

It is often in our opinion more convenient to use pressure (or stress) than density (or strain) as a control
parameter (a point we discuss below in Section~\ref{sec:asf}). 
We therefore use a stress-controlled procedure in our simulations, which is adapted from the Parrinello-Rahman
molecular dynamics (MD) scheme~\cite{PARA81}. 
The simulation cell has a 
rectangular parallelipipedic shape with lengths $L_\alpha$ parallel to coordinate axes $\alpha$ ($1\le \alpha\le 3$). 
$L_\alpha$ values
might vary, so that the system has $6N+3$ configurational degrees of freedom,which are the
 positions and orientations of the $N$ particles and lengths $L_\alpha$. $\Omega=L_1L_2L_3$ denotes the sample volume. 
We seek equilibrium states with set values $(\Sigma_{\alpha})_{1\le \alpha\le 3}$ of all three principal stresses 
$\sigma_{\alpha\alpha}$. 
We use the convention that compressive stresses are positive.

It is convenient to write position vectors ${\bf r}_i$, defining a square $3\times 3$ 
matrix with $L_\alpha$'s on the diagonal, as
$$
(1\le i\le N) \ \ {\bf r}_i = \ww{L}\cdot {\bf s}_i,
$$
${\bf s}_i$ denoting corresponding vectors in a cubic box of unit edge length. 
In addition to particle angular and linear velocities, which read
$$
(1\le i\le N) \ \ {\bf v}_i = \ww{L}\cdot \dot {\bf s}_i + \ww{\dot L}\cdot {\bf s}_i ,
$$
one should evaluate time derivatives $\dot L_i$. Equations of motion are written for particles in the standard form, 
\emph{i.e.,} (${\bf F}_i$ denoting the total force exerted on grain $i$) 
\be
(1\le i\le N) \ \ m_i \ddot {\bf s}_i = \ww{L}^{-1}\cdot {\bf F}_i,
\label{eqn:newtong}
\ee
and the usual equation for angular momentum. Meanwhile, lengths
$\dot L_\alpha$ satisfy the following equation of motion, in which ${\bf r}_{ij}$ is the vector joining the center of
$i$ to the center of $j$, subject to the
usual nearest image convention of periodic boundary conditions:
\be
\ba
M \ddot L_\alpha &=  \frac{1}{L_\alpha}\left[  L_\alpha^2 \sum_i m_i \left(\dot s_i^{(\alpha)}\right)^2 
+\sum_{i<j} F_{ij}^{(\alpha)} r_{ij} ^{(\alpha)} \right] \\
&- \frac{\Omega }{L_\alpha} \Sigma_\alpha.
\ea
\label{eqn:newtonc}
\ee
Within square brackets on the right-hand side of Eqn.~\ref{eqn:newtonc}, one recognizes the familiar 
formula~\cite{CMNN81,HaMcD86,BR90} for $\Omega \sigma_{\alpha\alpha}$, 
$\ww{\sigma}$ being the average stress in the sample:
\be
\sigma_{\alpha\beta} = \frac{1}{\Omega}\left[\sum_i m_i v_i^\alpha v_i^\beta +
\sum_{i<j} F_{ij}^{(\alpha)} r_{ij} ^{(\alpha)}
\right]
\label{eqn:stress}
\ee
 All three diagonal stress components should thus equate
the prescribed values $\Sigma_\alpha$ at equilibrium. The acceleration term will cause the cell to expand
 in the corresponding direction if the stress is too high, and to
shrink if it is too low. Eqn.~\ref{eqn:newtonc} involves a generalized mass $M$  
associated with the changes of shape of the simulation cell.
$M$ is set to a value of the order of the total mass of all particles in the sample.
This choice was observed to result in collective
degrees of freedom $L_\alpha$ approaching their
equilibrium values under prescribed stress $\ww{\Sigma}$ somewhat more slowly (but not exceedingly so) than
(rescaled) positions ${\bf s}$.

The original Parrinello-Rahman method was designed for conservative molecular
systems, in such a way that the set of equations is cast in Lagrangian form. This implies in particular additional
terms in~\eqref{eqn:newtong}, involving $\ww{\dot L}$. Such terms were observed to have a negligible
influence on our calculations and were consequently omitted. Granular materials are dissipative, 
and energy conservation is not an issue (except for some elastic properties, see paper III~\cite{iviso3}).
Further discussion of the stress-controlled method is provided in Appendix~\ref{sec:appendixcellm}. 

Equations~\eqref{eqn:newtong} and \eqref{eqn:newtonc}, with global degrees
of freedom $L_\alpha$ slower than particle
positions, lead to dynamics similar to those of a commonly used procedure in granular simulation~\cite{CUN88}.
This method consists in repeatedly changing 
the dimensions of the cell by very small amounts, then computing the motion of the grains
for some interval of time. A ``servo mechanism'' can be used
to impose stresses rather than strains~\cite{Makse04}. Our approach might represent a simplification,
as it avoids such a two-stage procedure.
It should be
kept in mind that we restricted our use of equation \eqref{eqn:newtonc}
to situations when changes in the dimensions of the simulation cell are very slow and gradual.
The perturbation introduced in the motion of the grains, in comparison to the more familiar case
of a fixed container, is very small. 

\subsection{Rigidity and stiffness matrices\label{sec:rist}}
We introduce here the appropriate formalism and state the relevant
properties of static contact networks. It is implied throughout this section that small displacements about an
equilibrium configurations are dealt with to first order (as an infinitesimal motion,
\emph{i.e.} just like velocities), and
related to small increments of applied forces, moments and stresses. In the following we shall exploit the definitions of stiffness matrices 
$\stia$ (Eqn.~\ref{eqn:sti1}) and $\stib$ to discuss stability properties of packings. The corrections to the degree of
force indeterminacy due
to free mechanism motions, as expressed by relations~\eqref{eqn:relhk} or~\eqref{eqn:relhk0}, will also be used.

The properties are stated in a suitable form
to the periodic boundary conditions with controlled diagonal stress components, as used in our
numerical study.
\subsubsection{Definition of stiffness matrix}
We consider a given configuration with bead center
positions (${\bf r}_i$, $1\le i\le n$) and orientations (${\bf \theta}_i$, $1\le i\le n$), and cell dimensions 
($L_\alpha$,  $1\le \alpha\le 3$). The grain center displacements $({\bf u}_i)_{1\le i\le n}$ are
conveniently written as 
$$
{\bf u}_i = \tilde {\bf u}_i - \ww{\epsilon}\cdot {\bf r}_i,
$$
with a set of displacements $\tilde {\bf u}_i$ satisfying periodic boundary conditions in the cell with the current 
dimensions, and the elements of the diagonal strain matrix $ \ww{\epsilon}$ express the relative shrinking deformation
along each direction, $\epsilon _\alpha = -\Delta L_\alpha /L_\alpha$. 
Gathering all coordinates of particle (periodic) displacements and rotation
increments, along with strain parameters, one defines a
\emph{displacement vector} in a space with dimension equal to the number of degrees of freedom $N_f=6n+3$,
\be
{\bf U} = \left( ( \tilde {\bf u}_i, \Delta {\bf \theta} _i)_{1\le i\le n},
(\epsilon_\alpha)_{1\le \alpha\le 3}\right).
\label{eqn:defu}
\ee
Let $N_c$ denote the number of intergranular contacts. In every contacting pair $i$-$j$,
we arbitrarily choose a ``first'' grain $i$ 
and a ``second'' one $j$. The normal unit vector $\nij$ points from $i$ to
$j$ (along the line joining centers for spheres).
The relative displacement $\delta {\bf u}_{ij}$ is defined for spherical grains with radius $R$ as
\be
\delta {\bf u}_{ij}= \tilde {\bf u}_i +\delta {\bf \theta}_i\times R \nij -\tilde {\bf u}_j
+\delta {\bf \theta}_j\times R \nij + \ww{\epsilon}\cdot {\bf r}_{ij}, \label{eqn:deprel1}
\ee
in which ${\bf r}_{ij}$ is the vector pointing from the center of the first sphere $i$ 
to the nearest image of the center of
the second one $j$. The normal part $\delta {\bf u}_{ij}^N$ of $\delta {\bf u}_{ij}$
is the increment of normal deflection $h_{ij}$ in the contact. 
\eqref{eqn:deprel1} defines a $3N_c\times N_f$ matrix $\ww{{\bf G}}$ 
which transforms ${\bf U}$ into the $3N_c$-dimensional
vector of relative displacements at contacts $\delta {\bf u}$:
\be
\delta {\bf u}=\rig \cdot {\bf U}
 \label{eqn:deprel}
\ee
In agreement with the literature
on rigidity theory of frameworks~\cite{BHSER98} ($-\ww{G}$ is 
termed normalized rigidity matrix in that reference), 
we call $\ww{{\bf G}}$ the \emph{rigidity matrix}. 

In each contact a force ${\bf F}_{ij}$ is transmitted from $i$ to $j$, which is split into its normal
and tangential components as ${\bf F}_{ij} = N_{ij}{\bf n}_{ij} + {\bf T}_{ij}$. 
The static contact law (without viscous terms) expressed in Eqns.~\eqref{eqn:hertz}, \eqref{eqn:tang}, with the
conditions stated in Section~\ref{sec:forces},
relates the $3N_c$-dimensional
contact force increment vector $\Delta {\bf f}$, formed with the values
$\Delta N_{ij}$, $\Delta \Tij $ of the normal and tangential parts of all contact force increments, to 
$\delta {\bf u}$:
\be
\Delta {\bf f} = \kk \cdot \delta {\bf u}.
\label{eqn:cmat}
\ee
This defines the $(3N_c\times 3N_c)$ matrix of contact stiffnesses $\kk $. $\kk $ is block diagonal (it does not
couple different contacts), and is conveniently written on using coordinates with $\nij$ as the first basis unit vector. 
In simple cases the $3\times 3$  block of $\kk$
corresponding to contact $i,j$, $\kk _{ij}$ is diagonal itself 
and contains stiffnesses $K_N(h_{ij})$ and (twice in 3 dimensions) $K_T(h_{ij})$ as
given by~\eqref{eqn:kn} and~\eqref{eqn:tang}:
\be
\kk _{ij}=\bma K_N(h_{ij}) & 0 & 0 \\ 0 &K_T(h_{ij})& 0 \\ 0&0&K_T(h_{ij})
\ema
.
\label{eqn:kiju}
\ee
More complicated non-diagonal forms of $\kk _{ij}$, which actually depend on the \emph{direction}
of the increments of relatives displacements in the contact, are found if friction is fully mobilized (which does not happen
in well-equilibrated configurations), or corresponding to
those small motions reducing the normal contact force. The effects of such terms is small, 
with our choice of parameters, and is discussed in paper III~\cite{iviso3}.  

External forces ${\bf F}_i$ and moments ${\bf \Gamma}_i$ (at the center) 
applied to the grains, and diagonal Cauchy stress components $\Sigma_\alpha$ can be gathered in one $N_f$-dimensional
\emph{load vector} $\fext$:
\be
\fext = \left(({\bf F}_i,{\bf \Gamma}_i)_{1\le i\le n},(\Omega \Sigma_\alpha)_{1\le\alpha\le 3} \right),
\label{eqn:deffext}
\ee
chosen such that the work in a small motion is equal to $\fext \cdot {\bf U}$.
The equilibrium equations -- the statements that contact forces ${\bf f}$ balance load $\fext$ --
is simply written with the tranposed rigidity matrix, as
\be
\fext = \rigt\cdot {\bf f}.
\label{eqn:trig}
\ee
This is of course easily checked on writing down all force and moment coordinates, as well as the equilibrium
form of stresses:
\be
\Omega \Sigma_\alpha = \sum _{i<j} F_{ij} ^\alpha  r_{ij}^\alpha.
\label{eqn:stressequil}
\ee
As an example, matrices $\rig$ and $\rigt$ were written down in~\cite{McGaHe05} in the simple case of one mobile disk with 2 contacts
with fixed objects in 2 dimensions, the authors referring to $-\rigt$ as the ``contact matrix''. The same definitions and matrices are used
in~\cite{McHe06} in the more general case of a packing of disks. 

Returning to the case of small displacements associated with a load \emph{increment} $\Delta \fext$, one may write,
to first order in ${\bf U}$,
\be
\Delta \fext = \sti \cdot {\bf U},
\label{eqn:defsti}
\ee
with a \emph{total stiffness matrix} $\sti$, comprising two parts, $\stia$ and $\stib$, which we shall respectively
refer to as the \emph{constitutive}
and \emph{geometric} stiffness matrices. $\stia$ results from
Eqns.~\ref{eqn:deprel}, \ref{eqn:cmat} and \ref{eqn:trig}:
\be
\stia = \rigt\cdot \kk \cdot \rig.
\label{eqn:sti1}
\ee
$\stib$ is due to the change of the geometry of the packing. 
Its elements (see Appendix~\ref{sec:appendixTrot}), relatively to to their counterparts
in  $\stia$, are of order $F/K_N R \sim h/R$, and therefore considerably smaller in all practical cases.
The constitutive stiffness matrix is also called ``dynamical matrix''~\cite{OSLN03,SRSvHvS05}. 
One advantage of decomposition \eqref{eqn:sti1} is to separate out the effects of the contact constitutive
law, contained in $\kk$ and those of the contact network, contained in $\rig$. $\rig$ is sensitive in general to
the orientations of normal unit vectors $\nij$ and to the ``branch vectors'' joining the grain centers to
contact positions -- which reduce to $R\nij$ for spheres of radius $R$. $\stib$, on the other hand,
unlike $\rig$, is sensitive to the curvature of grain surfaces at the contact point~\cite{KuCh06,Bagi07}.
\subsubsection{Properties of the rigidity matrix}
To the rigidity matrix are associated the concepts (familiar in structural mechanics) of force and velocity (or
displacement) indeterminacy, of relative displacement compatibility and of
static admissibility of contact forces. Definitions and properties stated in~\cite{JNR2000} for frictionless grains,
straightforwardly generalize to packings with friction.

The degree of displacement indeterminacy
(also called degree of hypostaticity~\cite{JNR2000}) is the dimension $k$ of the kernel of $\rig$, the
elements of which are displacements vectors ${\bf U}$ which do not create relative displacements in the contacts:
$\delta {\bf u}=0$. Such displacements are termed (first-order) mechanisms. Depending on boundary conditions, 
a grain packing might have a small number $k_0$ of ``trivial'' mechanisms, for which the whole system moves as
one rigid body. In our case, attributing common values of $\tilde {\bf u}$ to all grains
gives $k_0=3$ independent global rigid motions. 

The \emph{degree of force indeterminacy} $\hhh$ (also called
degree of hyperstaticity~\cite{JNR2000}) is the dimension of the kernel of $\rigt$, or the number of
independent self-balanced contact force vectors. If the coordinates of ${\bf f}$ are regarded as the unknowns in 
system of equations~\eqref{eqn:trig}, and if $\fext$ is supportable, then there exists a whole $\hhh$-dimensional affine
space of solutions. 

From elementary theorems in linear algebra one deduces a general
relation between $\hhh$ and $k$~\cite{JNR2000}
\be
N_f+\hhh= 3N_c+k.
\label{eqn:relhk}
\ee

An \emph{isostatic packing} is defined as one devoid of force and velocity indeterminacy (apart from trivial
mechanisms). Excluding trivial mechanisms (thus reducing $N_f$ to $N_f-k_0$), and loads that are not orthogonal
to them, one then has a square, invertible rigidity matrix. To any load corresponds a unique set of equilibrium
contact forces. To any vector of relative contact displacements corresponds a unique displacement vector.

With frictionless objects, in which contacts only carry normal forces, it is appropriate to use $N_c$-dimensional 
contact force and relative displacement vectors, containing only normal components, 
and to define the rigidity matrix accordingly~\cite{JNR2000}. Then \eqref{eqn:relhk} should be written as
\be
N_f+\hhh= N_c+k.
\label{eqn:relhk0}
\ee
In the case of frictionless spherical particles, all rotations are mechanisms, hence a
contribution of 3 to $k$. Thus one may in addition ignore all rotations, and subtract $3n$ both from $N_f$ and from $k$, 
so that \eqref{eqn:relhk0} is still valid.
In such a case, the rigidity matrix coincides (up to a sign convention and normalization of its elements)
with the one introduced in central-force networks, trusses and tensegrity structures~\cite{THDU98}. Donev
\emph{et al.}, in a recent publication on sphere packings~\cite{DTS05}, call rigidity matrix what we defined as its transpose $\rigt$.

\subsection{Control parameters\label{sec:control}}
The geometry and the mechanical properties of sphere packings under given pressure $P$ depends on
a small set of control parameters, which can be conveniently
defined in dimensionless form~\cite{CR03,RoCh05}. 

Such parameters include friction coefficient $\mu$ and viscous dissipation parameter $\zeta$,
which were introduced in Sec.~\ref{sec:forces}.

The elastic contact law introduces a \emph{dimensionless stiffness parameter} $\kappa$, which
we define as:
\be
\kappa= \left(\frac{\tilde E}{P}\right)^{2/3}.\label{eqn:defkappa}
\ee
Note that $\kappa$ does not depend on bead diameter $a$.
Under pressure $P$, the typical force in a contact is of order $Pa^2$. It corresponds to a normal
deflection $h$ such that $Pa^2 \sim \tilde E \sqrt{a}h^{3/2}$ due to the Hertz law~\eqref{eqn:hertz}. Therefore,
$\kappa$ sets the scale of the typical normal deflection $h$ in Hertzian contacts, as $h/a \sim 1/\kappa$. 

In the case of monodisperse sphere packings in equilibrium in uniform state of stress $\ww{\sigma}$,
pressure $P=\tr \ww{\sigma}/3$ is directly related to the average normal force $\ave{N}$.
Let us denote as $\Phi$ the solid fraction and $z$ the coordination number ($z=2N_c/n$). 
As a simple consequence of the classical formula
for stresses recalled in Sec.~\ref{sec:bcsc} (Eqn.~\ref{eqn:stress}
in the static case, or Eqn.~\ref{eqn:stressequil}), one has, neglecting contact deflections before diameter $a$,
\be
P= \frac{z\Phi \ave{N}}{\pi a^2},
\label{eqn:relpn}
\ee 
whence an exact relation between $P$ and contact deflections:
$$
\frac{\ave{h^{3/2}}}{a^{3/2}}=\frac{\pi}{z\Phi\kappa^{3/2}}.
$$
The limit of rigid grains is approached as $\kappa \to \infty$. $\kappa$ can reach very high values for
samples under their own weight, but most laboratory results correspond to levels of confining pressure 
in the 100~kPa range. Experimental data on the
mechanical properties of granular materials in quasistatic conditions 
below a few tens of kPa  are very scarce (see, however, \cite{Tat39} and \cite{ReCl01}). This is motivated by
engineering applications (100~kPa is the pressure below a few meters of earth), and this also results from
difficulties with low confining stresses. Below this pressure range, stress fields are no longer uniform, 
due to the influence of the sample weight, and measurements are difficult (\emph{e.g.}, elastic waves of measurable
amplitude are very strongly damped).

We set the lowest pressure level for our simulation of glass beads to 1~kPa or 10~kPa, which corresponds 
to $\kappa \simeq 181000$ and $\kappa \simeq 39000$. 
Such values, as we shall check, are high enough for some characteristic properties of rigid sphere
packings to be approached with good accuracy. Upon increasing $P$, 
the entire experimental pressure range will be explored in the two companion papers~\cite{iviso2,iviso3}.

Another parameter associated with contact elasticity is the ratio of tangential to normal stiffnesses (constant in our model),
related to the Poisson ratio of the material the grains are made of. Although we did not investigate the role
of this parameter, several numerical studies~\cite{Gael2,SRSvHvS05} showed its influence on global properties to be very small.

The ``mass'' $M$ of the global degrees of freedom is chosen to ensure slow and gradual changes in cell dimensions, and dynamical
effects are consequently assessed on comparing the strain rate $\dot \epsilon$ to intrinsic inertial times, such as the time needed for
a particle of mass $m$, initially at rest, 
accelerated by a typical force $Pa^2$, to move on a distance $a$. This leads to the definition of 
a \emph{dimensionless inertia parameter} :
\be
I=\dot \epsilon\sqrt{m/aP}.\label{eqn:defI}
\ee
The quasistatic limit can be defined as $I\to 0$.  
$I$ was successfully used as a control parameter in dense granular shear
flows~\cite{GdR04,Dacruz05,PouliquenPG}, which might be modelled on writing down the $I$
dependence of internal friction and density~\cite{JoFoPo05,JFP06}. 

The sensitivity to dynamical parameters 
$I$ and $\zeta$ should be larger in the assembling stage (as studied in the present paper) than in the subsequent 
isotropic compression of solid samples studied in paper II~\cite{iviso2}, for which one attempts to approach the
quasi-static limit.  In the following we will assess the
influence of parameters $\mu$, $\kappa$, $I$ and $\zeta$ on sample states and properties.

\section{Low-pressure isotropic states of frictionless packings. \label{sec:asf}}
\subsection{Motivations}
Numerical
samples are most often produced by compression of an initially loose configuration (a granular gas) in which the grains do not
touch. If the friction coefficient is set to zero at this stage,
one obtains dense samples, which depend very little on chosen mechanical parameters. These 
frictionless configurations are in a particular reference 
state which was recently investigated by several groups~\cite{OSLN03,DTS05}. 
We shall dwell on such an academic model as 
assemblies of rigid or slightly deformable \emph{frictionless} spheres in mechanical equilibrium for several reasons. 
First, we have to introduce various characterizations of the microstructure of sphere packings that will be useful in the presence of friction too. 
Then, such
systems possess rather specific properties, which are worth recalling in order to assess whether some of them could be of relevance
in the general case. 
Frictionless packings also represent, as we shall
explain, an interesting limit case. 
Finally, one of our objectives is to establish the basic uniqueness, in the statistical sense, of
the internal state of such packings under isotropic, uniform pressures, provided crystallization is thwarted by a fast enough
dissipation of kinetic energy. 
\subsection{Assembling procedures}
\subsubsection{Previous results}
Since we wish to discuss a uniqueness property, we shall compare our results to published ones whenever they are available.
Specifically, we shall repeatedly refer to the works of O'Hern, Silbert, Liu and Nagel~\cite{OSLN03}, and of Donev, Torquato and Stillinger~\cite{DTS05}, 
hereafter respectively abbreviated as OSLN and DTS. Both are numerical studies of frictionless sphere packings under isotropic pressures. 

OSLN use elastic spheres,
with either Hertzian or linear contact elasticity. They control the solid fraction $\Phi$, and record the pressure at equilibrium.
Their samples (from a few tens to about
1000 spheres) are requested to minimize elastic energy at constant density. For each one, pressure and elastic energy vanish below a certain threshold packing fraction
$\Phi_0$, which is identified to the classical random close packing density. Above $\Phi_0$, pressure and elastic constants are growing functions of density. OSLN
report several power law dependences of geometric and mechanical properties on $\Phi-\Phi_0$ which we shall partly review.

DTS differ in their approach, as unlike OSLN (and unlike us) they use strictly rigid spherical balls, and approach the density of
equilibrated rigid, frictionless sphere packing \emph{from below}. They use a variant of the classical (event-driven) hard-sphere
molecular dynamics method~\cite{alder59,HaMcD86}, in which sphere diameters are continuously growing, the Lubachevsky-Stillinger (LS) algorithm~\cite{LS90,LSP91},
to compress the samples. DTS's approximation of the strictly rigid sphere packing as the limit of a hard sphere glass with 
very narrow interstices (gaps) between colliding neighbors (contact forces in the static packing are then replaced by transfers of momentum between neighbors), 
and their resorting to linear optimization methods~\cite{JNR2000,DTSC04}, enable them to obtain very accurate results in samples of 1000 and 10000 beads. 

DTS expressed doubts as to whether numerical ``soft'' (elastic) sphere systems could approach the ideal rigid packing properties, and   
both groups differ in their actual definition of \emph{jamming} and on the relevance and definition of the \emph{random close packing} concept. 
Relying on our own simulation results, we shall briefly discuss those issues in the following.
\subsubsection{Frictionless samples obtained by MD\label{sec:MDres}}
Our numerical results on packings assembled without friction are based on five different configurations of $n=4000$ beads
prepared by compression of a granular gas without friction. 
First, spheres are placed on the sites of an FCC lattice at packing fraction $\Phi=0.45$ (below the freezing density,
$\Phi\simeq 0.49$~\cite{VCKB02}). Then they are set in motion with random velocities, and left to interact in collisions that preserve kinetic energy,
just like the molecules of the hard-sphere model fluid studied in liquid state theory~\cite{HaMcD86,VCKB02}. We use the
traditional event-driven method~\cite{alder59}, in a cubic cell of fixed size, 
until the initial crystalline arrangement has melted. Then, velocities are set to zero, and the molecular dynamics method
of  Section~\ref{sec:model} is implemented with an external pressure equal to
10~kPa for glass beads ($\kappa\simeq 39000$). Energy is dissipated thanks to viscous forces in contacts, and the packing
approaches an equilibrium state. Calculations are stopped when the net elastic force on each particle is below
$10^{-4}a^2P$, the elastic contributions to the stress components equal the prescribed value $P$ with relative error
smaller than  $10^{-4}$ and the kinetic energy per particle is below $10^{-8}Pa^3$. On setting all velocities to zero, it is observed
that the sample does not regain kinetic energy beyond that value, while the unbalanced force level does not increase. We have thus
a stable equilibrium state. This is further confirmed by the absence of mechanism in  the force-carrying contact network, apart from
the trivial free translational motion of the whole set of grains as one rigid body. From \eqref{eqn:sti1}, mechanisms coincide
with ``floppy modes'' of the constitutive stiffness matrix $\stia$ (\emph{i.e.}, the elements of its kernel). The geometric stiffness $\stib$, as
checked in Appendix~\ref{sec:appendixTrot}, is a very small correction (compared to those of $\stia$ the elements of
matrix $\stib$ are of order $\kappa^{-1}$).

In the following, such configurations assembled without friction will be referred to as A states. 

In order to check for a possible influence of the assembling procedure on the final configurations, we simulated
another, similar sample series, denoted as A', for which the LS algorithm was used to bring the solid fraction from $0.45$ to
$0.61$, before equilibrating at the desired pressure with Hertzian sphere molecular dynamics. 

Observed geometric and mechanical characteristics of A and A' states are reported below and compared to other published results, in
particuler those of OSLN and DTS. We also state specific properties of \emph{rigid} frictionless sphere packings, to which A configurations at high $\kappa$ 
are close. Unlike OSLN, we use pressure or stiffness level $\kappa$ as the control parameter. The state 
OSLN refer to as ``point J'', which appears as a rigidity threshold $\Phi=\Phi_0$ if solid fraction 
$\Phi$ is used as the control parameter, is approached here as $\kappa \to \infty$.

\subsubsection{Compression rates and duration of agitation stage}
Molecular dynamics is not the fastest conceivable route to minimize the sum of elastic and potential energies, and the MD approach does not necessarily
find the nearest minimum in configuration space. For that purpose, 
the direct conjugate gradient minimization approach, as used by OSLN, which involves no inertia and 
follows a path of strictly decreasing energy in configuration space is the best candidate.
 
However, the time scales
involved in the MD simulations can be compared to experimental ones.
In simulations, A configurations approach their
final density within a few tens of time $\tau = \sqrt{m/(aP)}$, and come to their final equilibrium with a few hundreds of $\tau$. 
Comparable laboratory experiments in which dense samples are assembled are sample preparation with the pluviation or rain deposition technique, 
in which grains are deposited at constant flow rate under gravity, with a constant height of 
free fall~\cite{RT87,ERCCD05,BENA01,BeCaDu04}. Such an assembling technique
produces homogeneous samples. Grains are first agitated near the free surface, and then subjected to a quasistatic pressure increase as pouring procedes. 
The relevant pressure scale corresponds therefore to the weight of the agitated superficial layer of
the sample  being assembled~\cite{ERCCD05},
typically of the order of 10 diameters, hence $P\sim 10 mg/a^2 $ and $\tau \sim \sqrt{a/(10g)}$, about $3\times 10^{-3}$~s for $a=1$~mm. 
Approximating the compaction time by the time needed to renew entirely the agitated superficial layer, 
we obtain a few times $10^{-2}$~s if this time is to be of order
$10 \tau$, as in our simulations. This corresponds to a fraction of a second to fill up a $10$~cm high container, 
a value within the experimental range. The main conclusion
from this crude analysis is that laboratory assembling processes are rather fast, 
with typical compaction times similar to those of our simulated isotropic 
compression procedure.

On the other hand, the LS procedure followed by DTS, which we used to produce our A' samples, 
unavoidably involves many collisions and a considerable level of agitation
while particle diameters grow at a prescribed rate. 
In practice, kinetic energy actually increases on implementing the LS algorithm: receding velocities after a collision have to
be artificially increased in order to make sure particles that are continuously growing in diameter actually 
move apart after colliding~\cite{LS90,LSP91}. Velocities have to be scaled down 
now and then for computational convenience, a feature the actual compacting process, 
depending on the ratio of growing rate to quadratic velocity average, is sensitive to.
In our implementation there were typically 110 collisions per sphere in the range $0.49\le \Phi \le 0.58$ (the most dangerous interval
for crystal nucleation~\cite{VCKB02,LGROT02}), and 90 collisions for $0.58\le \Phi \le 0.61$. 
DTS report using expansion rates of $10^{-4}$, while ours started as $10^{-2}$, in units of the quadratic mean velocity. 

Consequently, the order of simulation results, from the fastest, 
least agitated case to the slowest one is as follows: first the OSLN results, then our A,
followed by our A' series, and finally the simulations by DTS (who used a slower implementation of the LS method than our A' one).

\subsection{Energy minimization and density\label{sec:phimax}}
\subsubsection{What is ``jamming'' ?\label{sec:jam}}
In spite of a long tradition of studies on the geometry of sphere assemblies, the connection between mechanical equilibrium
and density maximization has seldom been stressed. This property was presented, in slightly different forms, in the mathematics~\cite{CO88}
and physics~\cite{JNR2000,DTS05} literature. 
It is worth recalling it here, as the purpose of this work is to discuss
both geometric and mechanical properties of such particle packings. This connection is simply expressed on noting that 
configurations of rigid, frictionless, non-adhesive spherical particles 
in stable equilibrium under an isotropic confining pressure are those that realize 
a local minimum of volume in configuration space, under the constraint of mutual impenetrability. 
It is no wonder then that the isotropic compaction of frictionless balls is 
often used as a route to obtain dense samples~\cite{TH00,Makse04}. 
In DTS~\cite{DTS05} and in other works by the same group~\cite{KTS02,DTSC04}, the authors use a definition of \emph{strictly jammed} 
configurations of hard particles as those for which particles
cannot move without interpenetrating or increasing the volume of the whole system. Their definition is therefore exactly equivalent
to that of a stable equilibrium state with rigid, frictionless grains under an isotropic confining pressure.

If we now turn to elastic, rather than rigid, spherical particles, with Hertzian contacts as defined in Sec.~\ref{sec:model}, then
stable equilibrium states under given pressure $P$ 
are local minima of the potential energy defined as ($H$ denotes the Heavside step
function)
\be
W = P\Omega + \sum _{1\le i<j\le n} \frac{2\tilde E\sqrt{a}}{15} h_{ij} ^{5/2} H(h_{ij}).
\label{eqn:defWe}
\ee
As stiffness parameter $\kappa$ increases, the second term of~\eqref{eqn:defWe} imparts an increasing
energetic cost to elastic deflections $h_{ij}$, and the solution becomes an approximation to a minimum of the first term, 
with impenetrability constraints, \emph{i.e.,} a stable equilibrium state of rigid, frictionless balls. The value of $\kappa$
is an indicator of the distance to the ideal, rigid particle configuration, and it is arguably more convenient to use that the
density, used by OSLN, because it does not vary between different samples.  
OSLN had to adjust the density separately for each sample in order to approach the limit
of rigid grains, so that the pressure approached zero, corresponding to a rigidity threshold. Their definition of jamming is based on
a local minimum of elastic energy, and therefore also coincides with ours: a jammed state is a stable equilibrium state.
\subsubsection{Solid fractions}
\begin{figure}[htb]
\includegraphics*[width=8.5cm]{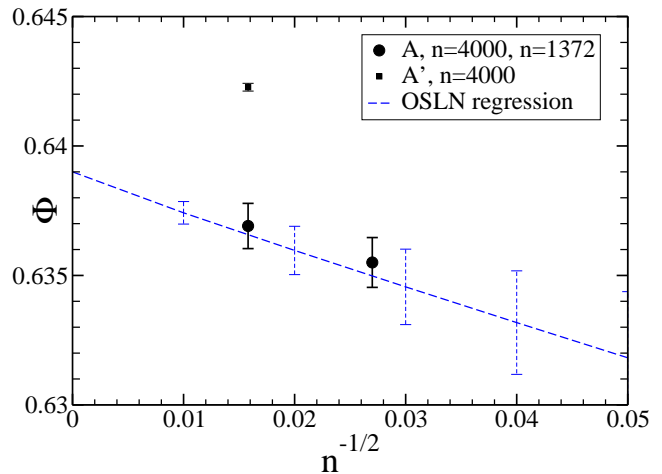}
\caption{\label{fig:phi}
(Color online) Solid fraction $\Phi$ versus $n^{-1/2}$. Blue dotted line: average value, with standard deviation indicated as error bars, 
according to OSLN's results, extrapolated to $n\to \infty$. Black round dots with error bars:
our A samples for $n=4000$ and other, very similar results for $n=1372$. Square dot:  A' samples with $n=4000$ (this point has
a smaller error bar). 
}
\end{figure}

Our A configurations have a solid fraction $\Phi = 0.6370 \pm 0.0015$ (indicated error bars correspond throughout the paper to
one sample-to-sample standard deviation). We shall check below that the small
density difference between $\kappa = 39000$ and $\kappa \to \infty$ is much smaller than the statistical uncertainty on $\Phi$.
OSLN performed a careful statistical analysis of finite size effects and uncertainty on $\Phi$, leading to estimates shown on Fig.~\ref{fig:phi}.
Fig.~\ref{fig:phi} also shows another MD data point we obtained for $n=1372$. Our $\Phi$ values
coincide with OSLN's estimation of size-dependent averages and fluctuations, once it is extrapolated to larger sample sizes (or, possibly, our configurations are
 very slightly denser). Our A' samples exhibit higher densities than A ones, $\Phi = 0.6422 \pm 0.0002$ -- a fairly small difference, but clearly larger than
error bars. 
DTS do not report $\Phi$ values very precisely, but 
mention solid fractions in the range 0.625 to 0.63~\cite[page 7]{DTS05}, on excluding the volume of rattlers, particles that 
transmit no force. This entails $0.639\le \Phi\le 0.644$ once those inactive grains, which represent about 2.2\% of the total number, 
are taken into account. 
LS-made samples were shown
in~\cite{TTD00} to jam, depending on the compression rate, over the whole solid fraction range between 0.64 and the
maximum value $\pi/(3\sqrt{2})$ corresponding to the perfect FCC crystal.
The final values of the solid fraction therefore correlate with the duration of the agitation stage in the assembling procedure. The RCP density is traditionnally
associated with a minimization of crystalline order. In the next section we check for indications of
incipient crystalline order in A and A' samples.
\subsection{Traces of crystalline order\label{sec:cryst}}
The possible presence of crystal nuclei, 
the FCC and HCP lattices (the former the more stable thermodynamically) and hybrids thereof
being the densest possible arrangements, is a recurring issue in sphere packing studies. A recent numerical study of crystallization dynamics in the hard sphere 
fluid is that of Volkov~\emph{et al.}~\cite{VCKB02}, in which the
authors used several indicators and measures of incipient crystalline order that we apply here to A and A' states. 
First, \emph{bonds} are defined as (fictitious) junctions
between the centers of neighboring spheres if their distance is smaller than some threshold, often chosen as 
corresponding to the first minimum in pair corelation function $g(r)$ (about $1.4a$ in our case, see Sec.~\ref{sec:grsf}). 
Then, a \emph{local} order parameter is associated
to each grain $i$, as:
\be
Q_l ^{\text{loc}}(i)=\left[ \frac{4\pi}{2l+1}\sum_{m=-l}^{m=l}\vert \hat q_{lm}(i) \vert^2\right]^{1/2},
\label{eqn:defqil}
\ee
in which $\hat q_{lm}(i)$ is an average over all neighbors $j$ of $i$ numbered from $1$ to $N_b(i)$, the number of bonds of $i$:  
\be
\hat q_{lm}(i) = \frac{1}{N_b(i)}\sum _{j=1}^{N_b(i)}Y_{lm}({\bf n}_{ij}),
\label{eqn:defqi}
\ee
${\bf n}_{ij}$ denoting as usual the unit vector pointing from the center of $i$ to the center of $j$.

$Q_4$ and $Q_6$, in particular, have been used to distinguish different local orders~\cite{VCKB02,AsSaSe05,LGROT02}.
In the sequel we use the average $Q_6^{\text{loc}}$ of \eqref{eqn:defqi} over all grains, as well as a global parameter $Q_6$, defined on taking the average
over all bonds within the
sample, instead of those of a particular grain $i$ in~\eqref{eqn:defqil}. The values of those parameters are given in table~\ref{tab:aacryst}.  
Global $Q_6$ values are small in large samples, because they tend to average to zero in the presence of randomly oriented polycrystalline textures. They can be used 
nevertheless to observe crystallization in samples of $\sim 10000$ beads, as they finally reach values comparable to the perfect crystal one~\cite{LGROT02}. 
 
Next, following~\cite{VCKB02}, we normalize the set $(\hat q_{lm}(i))_{-l\le m\le l}$, on multiplying, for any given $l$ and $i$, 
each of its $2l+1$ components by an appropriate common factor thus obtaining $(\overline q_{lm}(i))_{-l\le m\le l}$,
such that 
\be
\sum_{m=-l}^{m=l} \vert \overline q_{lm}(i) \vert ^2 =1. 
\label{eqn:defqin}
\ee
If the values $\hat q_{lm}(i)$ are viewed as the components of a  $2l+1$-dimensional local order parameter, then $\overline q_{lm}(i)$ might be viewed as its ``phase'' or ``angular''
part, characteristic of the choice of a direction, rather than of the intensity or extent 
with which the system is locally ordered. Then a bond is termed \emph{crystalline} if it
joins two particles for which those ``phases'' are sufficiently correlated: (the star indicates complex conjugation)
\be
\left\vert \sum_{m=-l}^{m=l}  \overline q_{lm}(i)   \overline q_{lm}^*(j) \right\vert \ge 0.5. 
\label{eqn:q6corrij}
\ee
A particle is said to be in a \emph{crystalline configuration} if at least 7 of its bonds 
(out of 12.5-13, see table~\ref{tab:aacryst})) is ``crystalline'', 
according to definition \eqref{eqn:q6corrij} with $l=6$. One may check how numerous those particles
are and whether they tend to cluster in crystalline regions. Table~\ref{tab:aacryst} contains those various indicators, 
as observed in samples of type A and A' at the largest
studied stiffness level. Order parameters have a very small value, indicating as expected a large distance to crystal order.
Only a small fraction of bonds and grains are declared ``crystalline'' according to the above definitions. However, it does transpire from the
data of table~\ref{tab:aacryst} that A' states are consistently more ``ordered'' than A ones, with a small, but systematic difference for all listed indicators
(see also Appendix~\ref{sec:appaste}).
\begin{table*}
\centering
\begin{tabular}{|cc|ccccccc|}  \cline{1-9}
State& $d_c/a$& $Z$ & $Z_{cr}$ & $Q_6$ &  $Q_6^{\text{loc}}$ &$Q_6^{\text{loc,cr}}$ &$x_{cr}$ & $\ave{n_{cr}}$\\
\hline
\hline
A & $1.35$& $12.36\pm 0.03$ &$2.91\pm 0.06 $&$(1.7\pm 0.3)\times 10^{-2}$
&$0.392 \pm 0.001$ & $0.417 \pm 0.003$ & $0.080 \pm 0.005$ & $19.8$\\
\hline
A'  & $1.35$&$12.50\pm 0.02$ &$3.13\pm 0.11 $&$(1.9\pm 0.5)\times 10^{-2}$
&$0.398 \pm 0.0005$&$0.420 \pm 0.002$& $0.104 \pm 0.006$& $54.8$\\
\hline
A  &$1.40$&$13.11\pm 0.02$ &$2.94\pm 0.06 $&$(1.6\pm 0.3)\times 10^{-2}$
& $0.370 \pm 0.001$ & $0.394 \pm 0.002$ & $0.083 \pm 0.006$ & $22.8$\\
\hline
A'  &$1.40$& $13.20\pm 0.02$ &$3.16\pm 0.11 $&$(1.8\pm 0.3)\times 10^{-2}$
&$0.377 \pm 0.0006$&$0.397 \pm 0.003$& $0.103 \pm 0.006$& $64.5$\\
\hline
\end{tabular}
\normalsize
\caption{Indicators of possible incipient crystalline order in states A and A' at $\kappa=39000$. $Z$ is the coordination number of first heighbors, $Z_{cr}$ the
``crystalline bond'' coordination number,  $Q_6$ and  $Q_6^{\text{loc}}$ the global and (average) local order parameters, $Q_6^{\text{loc,cr}}$
its average value within crystalline regions, $x_{cr}$ the fraction of ``crystalline'' particles and
$\ave{n_{cr}}$ the mass average of the  number of particles in a ``crystalline cluster''. First neighbors are defined here as those closer than distance $d_c=1.35a$ or $d_c=1.40a$, near
the first minimum in $g(r)$.
\label{tab:aacryst}}
\end{table*}
Most notable is the increase of the size of ``crystalline'' regions.
A direct visualization of those domains, as we checked, shows that they are quite far from perfectly ordered, but reveals
some local tendency to organization in parallel stacked layers, and to the formation of 2D triangular lattice patterns within the layers.
Luchnikov \emph{et al.}~\cite{LGROT02} report simulation of 16000 particle samples of the hard sphere fluid 
evolving towards crystallization at constant density (between $\Phi=0.55$ and $\Phi=0.6$), as monitored by
the global $Q_6$ parameter and the distribution of local $Q_6$ values. They observed, like Volkov \emph{et al.}~\cite{VCKB02}, that
several thousands of collisions per particle were necessary for a significant evolution to take place, which is compatible with our observation of 
a detectable, but very small tendency with about 100 collisions per particle with our A' samples. 

$Q_6^{\text{loc}}$ and $Q_4^{\text{loc}}$, as defined in~\eqref{eqn:defqil}, were also used by Aste \emph{et al.}~\cite{AsSaSe05} to characterize local arrangements, 
in an experimental study of sphere packing geometry by X-ray tomography. These results rely on observations of large samples of tens to hundreds of
thousands of beads, although not isotropic. Particles are classified according to the pair of values $Q_6^{\text{loc}}(i)$, $Q_4^{\text{loc}}(i)$. 
We compared the geometry of our numerical samples of similar density to those experimental data, 
with the result that although the most frequently observed values of
 $Q_6^{\text{loc}}(i)$ and $Q_4^{\text{loc}}(i)$ were quite close 
to experimental ones in dense samples, and the proportion of hcp-like particles were similar, fcc-like local environments were exceptional in simulations, whereas
a few percent of the spheres were classified in that category in the experimental results. Quantitative results are given in Appendix~\ref{sec:appaste}. 
Nucleation of crystalline order is strongly sensitive to sample history and boundary conditions~\cite{VCKB02}. 
\subsection{Properties of force networks\label{sec:asfbb}}
\subsubsection{Identification and treatment of rattlers\label{sec:ratt}}
The \emph{rattlers} are defined as the grains that do not participate
in carrying forces and remain, therefore,
free to ``rattle'' within the cage formed by their force-carrying, rigidly fixed neighbors. 
We refer to the network of contacting grains that carry forces as the \emph{backbone}. The backbone is the structure formed by non-rattler grains.
The fraction $x_0$ of rattlers  at $\kappa = 39000$ is $x_0= 0.013\pm 0.002$ in A samples, and it is slightly higher, $x_0=0.018\pm 0.002$ in A' ones.
DTS report $x_0\simeq 0.022$, and hence once again our A' results are closer to theirs. 
The proportion of rattlers increases slightly for stiffer contacts (higher $\kappa$ values).

Distinguishing between the backbone and the rattlers requires some care, as very small forces on the backbone might be confused with forces below tolerance
between rattler and backbone grains, or between two rattlers. We apply the following simple procedure. First, we regard as rattlers all spheres having less than
four contacts: less than three contacts implies a mechanism, and only three is impossible if forces are all strictly compressive. We
also discard from the backbone all spheres with only forces smaller than the tolerance. Then, all the contacts of eliminated spheres being also removed, other 
spheres might (although this is an extremely rare occurrence) have less than four contacts, so the procedure is iterated (twice at most
is enough in our samples, although one such sweep is usually enough) 
until no more rattler is detected. We found this method to work correctly 
for $n=4000$ and $\kappa=39000$. If one eliminates too many particles, the identified backbone might become floppy (hypostatic). 
We check, however, that its constitutive stiffness matrix remains positive definite, thereby avoiding such pitfall. 
The proportion of rattlers is likely to increase for stiffer contacts (higher $\kappa$). 

The presence of rattlers complicates the analysis of geometric properties of static packings, because their positions are not determined by the equilibrium
requirement. The rattlers are free to move within a ``cage'' formed by their backbone neighbors, 
and there is no obvious way in principle to prefer one or another of their infinitely many 
possible positions. This renders the evaluation of geometric data like pair correlations somewhat ambiguous. Moreover, rattlers, although scarce in
frictionless packings, can be considerably more numerous in frictional ones (see Section~\ref{sec:assemblef}). We therefore specify whether the results
correspond to direct measurements on the configurations resulting from the simulations, 
with rattlers floating in some positions resulting from compaction
dynamics, or whether rattlers have been fixed, each one having three contacts with the backbone (or some previously fixed other rattler). To compute such
fixed rattler positions with MD, we regard each backbone grain as a fixed object, exert small isotropically distributed random forces on all rattlers and let them move
to a final equilibrium position (assuming frictionless contacts). A third possibility is to eliminate rattlers altogether before recording geometric data. 
These are three choices
referred to as I, II and III in the sequel, and we denote observed quantities with superscripts I, II or III accordingly. 

Packings under gravity, if locally in an isotropic
state of stress, are expected to be in the same internal state and to exhibit the same properties as the ones that are simulated here. In such a situation, individual
grain weights are locally, within an approximately homogeneous subsystem, dominated by the externally imposed isotropic pressure. 
There is no rattler under gravity, but some grains
are simply feeling their own weight, or perhaps that of one or a few other grains relying on them. 
Such grains are those that would be rattlers in the absence of gravity. 
Instead of
freely floating within the cage of their backbone neighbors, they are supported by the cage floor. 
The situation should therefore be similar to that of our samples after all rattlers
have been put in contact with the force-carrying structure (treatment II), except that the small external forces applied to them are all directed downwards. 
\subsubsection{Coordination numbers}
Table~\ref{tab:coord} gives the distribution of local coordination number values among the spheres for A and A' states at $\kappa=39000$. In this table, $x_i$ is the
proportion of grains with $i$ contacts. 
\begin{table}[htb!]
\centering
\begin{tabular}{|c|cccccccccccc|}  \cline{1-13}
State& $x_0$ & $x_1$& $x_2$& $x_3$& $x_4$& $x_5$& $x_6$& $x_7$& $x_8$& $x_9$& $x_{10}$& $x_{11}$\\
\hline
\hline
A (I) & $1.3$ & $0$& $0$& $0$& $11.1$& $23.2$& $28.4$& $22.6$& $10.3$& $2.8$& $0.3$& $0.02$\\
\hline
A' (I)& $1.8$ & $0$& $0$& $0$& $11.6$& $22.5$& $28.2$& $22.3$& $10.6$& $2.8$& $0.3$& $0.01$\\
\hline
A (II) & $0$ & $0$& $0$& $1.1$& $10.7$& $22.7$& $28.2$& $22.9$& $10.9$& $3.1$& $0.3$& $0.02$\\
\hline
A' (II)& $0$ & $0$& $0$& $1.7$& $10.9$& $21.9$& $28.2$& $22.5$& $11.4$& $3.1$& $0.3$& $0.01$\\
\hline
\end{tabular}
\normalsize
\caption{Percentages $x_i$ of grains having $i$ contacts in A and A' configurations, on ignoring the rare contacts with or between
rattlers (I), or on fixing the rattlers onto the backbone with small (randomly oriented) forces (II).
\label{tab:coord}}
\end{table}
If rattlers are stuck to the backbone (method II),
one records slightly changed proportions of spheres with $n\ge 3$ contacts, to which values observed within samples
under gravity should be compared.
Distributions of local coordination numbers observed by DTS coincide to the data of Table~\ref{tab:coord} within 1\%. We attribute this small difference to the 
influence of contact deflections of order $\kappa^{-1}a$ in the MD results, 
while the DTS results are closer to ideally rigid packings (approached as open gaps tend to zero).
\subsubsection{Isostaticity\label{sec:asfiso}}
We now discuss how the \emph{isostaticity property}~\cite{OR97b,JNR97b,MO98a,MO98b,TW99,JNR2000,MO01,Mou04} 
of equilibrium states of rigid, frictionless spheres, influences
high $\kappa$ configurations of type A.

Isostaticity is a property of the \emph{backbone}, \emph{i.e.} the force-carrying contact network, in equilibrium packings of rigid, frictionless spheres.
It means that such networks are both devoid of hyperstaticity (force indeterminacy) and of hypostaticity (displacement indeterminacy), 
apart from possible trivial motions in which
all force-carrying grains move as one rigid body. These two properties have different origins~\cite{JNR2000}, 
and are not valid under the same assumptions.
The absence of hyperstaticity ($\hhh=0$ with the notations of Sec.~\ref{sec:rist}) 
results from the generic disorder of the packing geometry. It would hold true
for arbitrarily shaped rigid particles interacting by purely normal contact forces whatever the sign of those forces, and it applies to the whole packing, whatever
the contacts the rattlers might accidentally have with the backbone. The  
absence of hypostaticity property (except for trivial mechanisms, $k=k_0$), on the other hand,  
is only guaranteed for \emph{spherical} particles with \emph{compressive} forces in the contacts, and it applies to the sole backbone.

Due to the isostaticity property, the coordination number should be equal to 6 in the rigid limit \emph{on the backbone}. 
If $N_c$ is the number of \emph{force-carrying} contacts, then the
global (mechanical) coordination number is ${\disty z = \frac{2N_c}{n}}$ (possible contacts of the rattlers are discarded), and 
the backbone coordination number is defined as ${\disty z^* =  \frac{2N_c}{n(1-x_0)}=\frac{z}{1-x_0}}$. 
$z^*$, rather than $z$, has the limit $6$ as $\kappa\to +\infty$. In A samples ($\kappa=39000$) one has $z^* = 6.074 \pm 0.002$ (and hence
$z \simeq 5.995$), the excess over the limit $z^* =6$ resulting from
contacts that should open on further decreasing the pressure. 

The isostaticity property can be used to evaluate the density increase due to finite particle stiffness. To first order in the small displacements between $P=0$ 
(or $\kappa = +\infty$ ) and the current finite pressure state A, one might use the theorem of virtual work~\cite{JNR2000}, with the displacements
that bring all overlaps $h_{ij}$ to zero, and the current contact forces. Such motions leading to a simultaneous opening ($h_{ij}=0$) of all
contacts are only possible on networks with no hyperstaticity,
because there is no compatibility condition on relative normal displacements~\cite{JNR2000}.
This yields an estimate of the increase of the solid fraction $\Delta \Phi$ over its value
$\Phi_0$ in the rigid limit, as
\be
\frac{1}{\Omega}\sum _{ij} N_{ij} h_{ij} = P \frac{\Delta \Phi}{\Phi}.
\label{eqn:dphi1}
\ee
This equality can be rearranged using the Hertz contact law~\eqref{eqn:hertz} to relate  $N_{ij}$ to $h_{ij}$, and relation~\eqref{eqn:relpn}. We denote
as $Z(\alpha)$
the moment of order $\alpha$ of the distribution of normal forces $N_{ij}$, normalized by the average over all contacts:
\be
Z(\alpha) = \frac{\ave{N^\alpha}}{\ave{N}^\alpha}.
\label{eqn:defza}
\ee
\eqref{eqn:dphi1} can be rewritten as:
\be
\Delta \Phi = 3^{5/3}\Phi^{1/3} Z(5/3)\left(\frac{\pi}{z}\right)^{2/3} \kappa^{-1}.
\label{eqn:dphi}
\ee
In the isostatic limit which is approached at large $\kappa$, the force distribution and its moments are
determined by the network geometry, and we observed $Z(5/3)=1.284$. 
Taking for $z$ and $\Phi$ the values at the highest studied stiffness level $\kappa$ ($\kappa=39000$),  
this enables us to evaluate the density change between those configurations
and the rigid limit  as $\Delta \Phi \simeq 1.15\times 10^{-4}$. 
As announced before this is smaller than the statistical uncertainty on $\Phi$, and hence this does not
improve our estimation of the solid fraction $\Phi _0$ of the packing of rigid particles ($\kappa = +\infty$).
Recalling $\kappa^{-1} = (P/\tilde E)^{2/3}$, \eqref{eqn:dphi} means that the macroscopic relation between pressure and density has the same power law form
($P\propto (\Delta \Phi)^{3/2}$) as the contact law ($N \propto h^{3/2}$). This was observed by OSLN. It would hold, because of the isostaticity
property in the rigid limit, for whatever exponent $m$ in the contact law, the prefactor of the macroscopic relation $P\propto (\Delta \Phi)^m$ involving 
$Z(1+1/m)$, a moment of the geometrically determined force distribution.

As a consequence of~\eqref{eqn:dphi}, one can simply relate the \emph{bulk modulus} of frictionless packings to the pressure, 
as observed by OSLN too, a property which will be
used and discussed in paper III~\cite{iviso3}, which deals with elasticity of packings.
\subsubsection{Force distribution}
\begin{figure}[htb]
\includegraphics*[width=8.5cm]{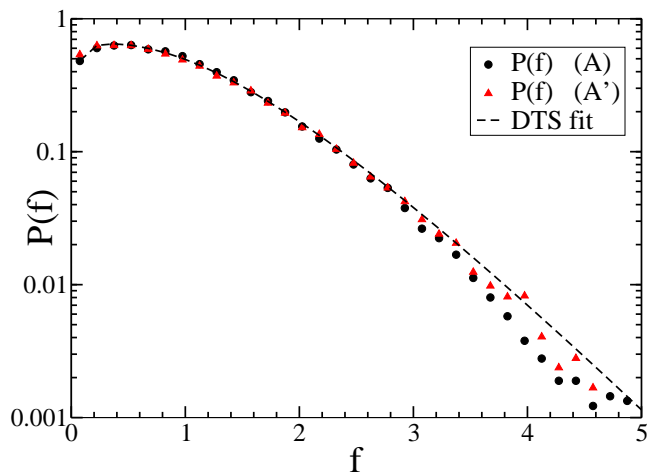}
\caption{\label{fig:frigsf} (Color online)
Probability distribution function $P(f)$ of normalized contact forces $f=N/\ave{N}$ in A and A' configurations at high $\kappa$. The dashed line is the fit proposed
by DTS: $P(f)=(3.43f^2+1.45-1.18/(1+4.71f))\exp(-2.25f)$.
}
\end{figure}

The force distribution we observe in A samples
at high $\kappa$ values approaches the one of a rigid packing, which due to isostaticity is a purely \emph{geometric} property. It is represented on 
Figure~\ref{fig:frigsf}. 
The data presented here are averaged over 5 samples. Because of relation~\eqref{eqn:relpn}, 
all samples prepared at the same pressure have the same average force, and this restores the ``self-averaging'' property, which OSLN observed  to be lacking on using
solid fraction instead of pressure as the control parameter. The choice of $\Phi$ as a state variable, because
of the finite size of the sample causing fluctuations of the threshold $\Phi_0$ where $P$ vanishes, is less convenient in that respect.

Fig.~\ref{fig:frigsf} also shows that the form proposed by DTS to fit their data is in very good agreement with our results, except perhaps
for large forces, for which it is a better fit for A' data -- thus providing additional evidence that A' samples are closer than A ones to the DTS results. 

\subsection{Geometric characterization\label{sec:asfgeom}}
\subsubsection{Pair correlation function\label{sec:grsf}}
\begin{figure}[htb]
\includegraphics*[width=8.5cm]{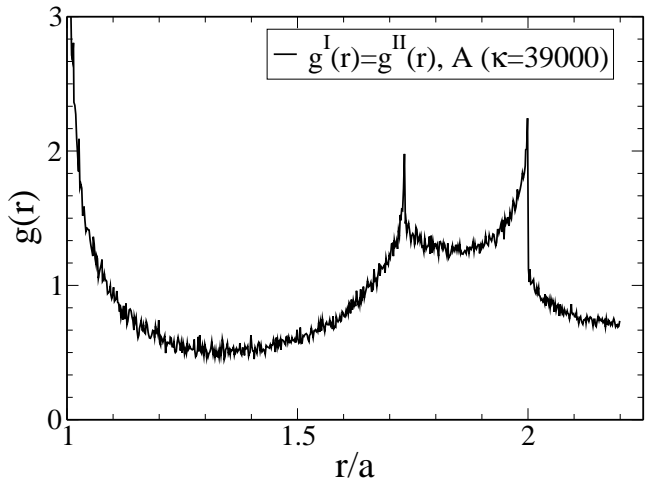}
\caption{\label{fig:gr}
Pair correlation functions $g^{I}(r)$ and $g^{II}(r)$ versus $r/a$ in A samples at $\kappa=39000$. Both definitions coincide on this scale (only
the peak for $r\to a^+$ is slightly different).
}
\end{figure}

Pair correlations should preferably 
be measured either with method I or method II, as there is no reason to eliminate rattlers before studying geometric properties. Comparisons
between pair correlation functions $g^{I}(r)$ and $g^{II}(r)$ (Fig.~\ref{fig:gr}) show very little difference on the scale of one particle diameter. 
Results of Fig.~\ref{fig:gr} 
are very similar to other published ones (\emph{e.g.,} in DTS), 
with an apparent divergence as $r\to a$ and a split second peak, with sharp maxima at $r=a \sqrt{3}$ and $r=2a$. 
The pair correlation function
should contain a Dirac mass at $r=a$ in the limit of rigid grains, which broadens into a sharp peak for finite contact stiffness. 
The weight of this Dirac term or peak in
the neighbor intercenter distance probability distribution function $4\pi \frac{n}{\Omega}r^2 g(r) = 24 (r^2/a^3)\Phi g(r)$ is coordination
number $z$, and the shape of the left shoulder of the peak at finite $\kappa$ is directly related to the force distribution $P(N)$:
$$\mbox{(For $\delta>0$)  }g(a-\delta) = \frac{za^3 \tilde E\sqrt{a\delta }}{48\Phi (a-\delta)^2}P(\frac{\tilde E\sqrt{a}\delta^{3/2}}{3}).$$
This explains the observation by OSLN~\cite{OSLN03} of the width of the $g(r)$ peak decreasing approximatively as $\Delta \Phi$, 
while its height increases as  $(\Delta \Phi)^{-1}$,
as the threshold density $\Phi_0$ is approached from above. 
The form of the distribution of contact forces, which is determined by the geometry of the isostatic backbone, remains exactly the same for all small enough
values of $\Delta \Phi$, with a scale factor proportional to $\Delta \Phi^{3/2}$, due to Eqns.~\ref{eqn:relpn} and \ref{eqn:dphi}. 

The sharp drop of $g(r)$ at $r=a\sqrt{3}$ and $r=2a$ was found by DTS to go to a discontinuity in the rigid particle limit. This can be understood as follows.
Each sphere has a number of first contact neighbors ($z$ on average) at distance $r=a$ if the grains are rigid, and a number of second contact neighbors
(\emph{i.e.} particles not in contact with it, but having a contact with at least one of its first contact neighbors). Such second contact neighbors will
make up for a significant fraction of particles with their centers at a distance $r\le 2a$, but none of them can be farther away. Futhermore, this leads to
a systematic depletion of the corona $2a<r<2a + \delta$ (with $0<\delta <a$) by steric exclusion. 
\subsubsection{Near neighbor correlations}
\begin{figure}[!htb]
\includegraphics*[angle=270,width=8.5cm]{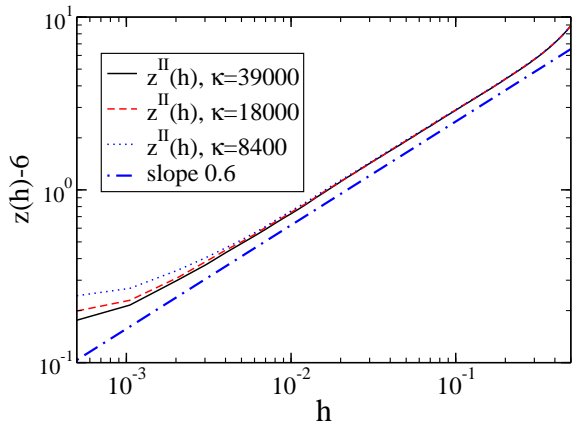}
\caption{(Color online)
Coordination number of near neighbors function of interstice $h$. The power law regime extends to smaller and smaller $h$ values as 
$\kappa$ increases. \label{fig:zh}
}
\end{figure}

As $r\to a^+$, pair correlations are conveniently expressed with the gap-dependent coordination number $z(h)$. 
$z(h)$ is the average number of neighbors of one sphere separated 
by an interstice narrower than $h$. $z(0)$ is the usual contact coordination number $z$. 
Function $z(h)$ has three possible different definitions
$z^{I}$, $z^{II}$ and $z^{III}$ according to the treatment of rattlers. All three of them 
were observed to grow as $z(0) + C h^{0.6}$ for $h$ smaller than about $0.3a$, constant $C$ taking slightly different values for $z^{I}$, $z^{II}$ and $z^{III}$.
$z^{III}(h)$ is equal to $z^*$ for $h=0$, and is very well fitted with the value $C=11$ found
by DTS~\cite[Fig. 8]{DTS05}. 
$z(h)$ deviates from this power law dependence corresponding to the rigid limit for small $h$, of the order of the typical overlap
$\kappa^{-1}$, as shown on Fig.~\ref{fig:zh}. This power law corresponds to $g(r)$ diverging as $(r-a)^{-0.4}$ as $r\to a^+$. 
Silbert \emph{et al.}, in a recent numerical study of states with high levels of rigidity~\cite{SiLiNa06} ($\kappa>10^6$), 
report observing $z^I(h)$ to grow with an exponent closer to $0.5$, although somewhat dependent on the choice of the $h$ interval for the fit.
However this does not contradict our main conclusion that different numerical approaches track the same
RCP state in the rigid limit.
\subsubsection{Other properties of contact networks}
Ignoring rattlers (method III) one may record the density of specific particle arrangements in the backbone. We thus find the contact
network (joining all centers of interacting particles by an edge) to comprise a number of equilateral triangles, such that on average each backbone grain belongs
to $2.04\pm 0.04$ triangles. In the rigid limit this gives a Dirac term for 
$\theta= \pi/3$ in the distribution of angles $\theta$ between pairs of contacts of the same grain. 
Tetrahedra are however very scarce (as observed by DTS), involving about $2.5\%$ of the beads, 
and pairs of tetrahedra with a common triangle are exceptional (5 such pairs in 5 samples
of 4000 beads). Pairs of triangles sharing a common base are present with a finite density, which explains the discontinuous drop at $r=a\sqrt{3}$ of $g(r)$,
this being the largest possible  distance for such a population of neighbor pairs.

\subsection{Conclusions\label{sec:asfdisc}}
We summarize here the essential results of Section~\ref{sec:asf}, about frictionless packings.
\subsubsection{Uniqueness of the RCP state}
Our numerical evidence makes a strong case in favor of the
uniqueness of the simulated rigid packing state made with frictionless spheres under isotropic pressure. Specifically, we observed quantitative agreements with
other published results~\cite{OSLN03,DTS05} in the coordination
numbers, the force distributions, the pair correlations and the frequency of occurrence of local contact patterns, 
even though different numerical methods have been used.
The small remaining differences in solid fraction, proportion of rattlers, 
and probability of large contact forces all correlate with the duration of agitated assembling stage, 
which can be measured in terms of numbers of collisions per grain at a given density. This duration directly correlates to the packing fraction and 
to the small amount of crystalline order in the samples. We therefore checked in an accurate, quantitative 
way the traditional views about \emph{random close packing} (RCP). The RCP
state can be defined
in practice as the unique state in which rigid frictionless spherical beads assemble 
in a static equilibrium state under isotropic pressure, in the limit of fast compression, so that
the slow evolution towards crystallization has a negligible influence. The Lubachevsky-Stillinger algorithm tends to produce packings
with a small but notable crystalline fraction. 
\subsubsection{Relevance of MD simulations, role of micromechanical parameters}
The uniqueness of the RCP implies that dynamical parameters $\zeta$ and $I$ have no influence on the frictionless packing structures, 
at least in the limit of fast compression rates.

Standard MD methods compare well with specifically designed methods that deal with rigid particles, and prove able to approach 
the rigid limit with satisfactory, if admittedly smaller, accuracy. 
Recalling that $\kappa = 39000$ corresponds to glass beads under $10$~kPa, it seems that laboratory samples under usual conditions
might in principle (if friction mobilization can be suppressed) approach the ideal (rigid particle) RCP state. 

Moreover, the time scales to assemble samples in MD simulations, if compared to
estimated  preparation times in the laboratory with such techniques as controlled pluviation, has the right order of magnitude. 
This means that the assembling proces is rather fast in experimental practice
when grains are deposited under gravity, which explains why densities above RCP are not directly obtained. 
Of course, in practice, many procedures produce anisotropic states.
Anisotropic packings of rigid, frictionless balls, under other confining stresses than a hydrostatic pressure,
should differ from the RCP state, and the numerical simulations of gravity deposited packings of frictionless beads of refs.~\cite{SEGHL02,SGL02} 
could be analysed in this respect. We chose here to study ideal preparation methods, and we only deal with isotropic systems. 
\subsubsection{Approach to isostaticity in the rigid limit}
We checked that bead packings under typical laboratory pressures such as 10~kPa might closely approach the isostaticity property of rigid frictionless packings.
We showed that some observations made by O'Hern~\emph{et al.}~\cite{OSLN03} 
on pressure or bulk modulus 
dependence on density, and on the shape of the first peak of $g(r)$ were direct consequences of this
remarkable property.
\section{Low-pressure states of frictional packings obtained by different procedures\label{sec:assemblef}}
\subsection{Introduction}
It is well known that the introduction of friction in granular packings tends in general to reduce density and coordination number, as observed
in many recent numerical simulations (see \emph{e.g.}~\cite{TH00,MJS00,SEGHL02,SGL02}), and that frictional granular assemblies,
unlike frictionless ones, can be prepared in quite a large variety of different states. In the field of soil mechanics, sand samples are traditionnally classified by
their density~\cite{DMWood,BiHi93,MIT93},
 which determines behaviors that have been observed in simulations of model systems as well~\cite{TH00,RR04}. (\emph{Inherent anisotropy}
of the fabric, \emph{i.,e.,}
the one due to the assembling process, rather than induced by anisotropic stresses, is a secondary, less influential state variable~\cite{Oda72,AM72}).
Engineering studies on sands usually resort to a conventional definition of minimum and maximum densities, based on standardized procedures~\cite{eminemax}. 

The motivation of the present study is to explore the range of accessible packing states, 
as obtained by different numerical procedures that produce homogeneous and isotropic periodic samples.
We therefore chose to bypass the painstaking computations needed to mimic actual laboratory assembling methods, 
but we argue that our procedures produce plausible structures with similar properties. 

One key result is that density alone does not determine the internal state of an isotropic packing, because the
coordination number can vary independently.

The A-type configurations obtained without friction in Section~\ref{sec:aaf} are local density maxima in configuration space (see Sec.~\ref{sec:jam}). 
Hence compaction
methods can be regarded as strategies to circumvent the mobilization of 
intergranular friction forces. Two such procedures are studied here, in a simplified, idealized form:
lubrication and vibration. We also simulate, as a reference, a state which can be regarded as a loose packing limit, at least with a definition relative to one
assembling method and friction coefficient $\mu=0.3$ ; and we prepare, as an interesting limit from a theoretical point of view, infinite friction samples.
 
Assembling procedures are described in Section~\ref{sec:aaf}, geometric aspects are studied in Section~\ref{sec:fgeom},
and contact network properties in Section~\ref{sec:ten}. Section~\ref{sec:aafdisc} summarizes the results. 
\subsection{Assembling processes for frictional grains\label{sec:aaf}}
Just like in the frictionless case, for each one of the packing states, we prepare 5 samples of 4000 beads, 
over which results are averaged, error bars corresponding to sample-to-sample fluctuations. The equilibrium criteria are those of Section~\ref{sec:MDres},
supplemented with a similar condition on moments. To identify rattlers, we use the procedure defined in Section~\ref{sec:ratt}, which is adapted to the
case of frictional grains: spheres with as few as
two contacts may carry forces (even large ones, as we shall see) and should not be regarded as rattlers.
\subsubsection{Looser packings compressed with final friction coefficient\label{sec:aaf1}}
We used direct compression of a granular gas in the presence of 
friction ($\mu=0.3$), another standard numerical procedure~\cite{MJS00,TH00,SUFL04}, in which the obtained density and coordination number are decreasing functions 
of $\mu$~\cite{TH00,MJS00,SEGHL02,SGL02}. This produces
rather loose samples hereafter referred to as D (B and C ones, to be defined further, denoting denser ones, closer
to A, but arguably more ``realistic''). D samples were made with exactly the same method as A ones (see Sec.~\ref{sec:MDres}), 
except that the friction coefficient  $\mu=0.3$ was used instead of $\mu=0$. 
In principle, D configurations should depend on initial compaction dynamics: increasing  the rate of compression could 
produce denser equilibrated packings, just like a larger height of free fall, whence a larger initial kinetic energy, increases the density of configurations obtained
by rain deposition under gravity~\cite{ERCCD05}. We request the reduced compression rate $I$, 
defined in~\eqref{eqn:defI}, not to exceed a prescribed maximum value $I_{\text{max}}$.
The choice of $I_{\text{max}}=10^{-3}$ and $\zeta=0.98$ yields solid fractions
$\Phi= 0.5923\pm 0.0006$ and backbone coordination numbers $z^* = 4.546 \pm 0.009$, with a rattler
fraction $x_0 = (11.1 \pm 0.4)\%$.
These data correspond to $P= 1$~kPa (or $\kappa \simeq 181000$). Very similar results are obtained on using a different, 
but low enough pressure, such as $10$ or even $100$~kPa, 
as remarked in~\cite{SRSvHvS05} (where 2D samples were assembled by
oedometric compression), and as indicated in table~\ref{tab:initP}. However, a quasistatic compression from $P= 1$~kPa to $10$ or $100$~kPa
produces slightly different states at the same pressure. 
The influence of $\zeta$ should  disappear in the limit of slow compression, $I\to 0$. 
A practical definition of a ($\mu$-dependent) limit of loose packing obtained by direct compression, 
can therefore be proposed as the $I\to 0$ limit of our D states.
\begin{table*}
\caption{Isotropic states of type D, from direct compression of the granular gas at the indicated pressure (rows marked ``gas''),
or from gradual, quasistatic compression (rows marked ``QS'') of solid samples made at the lowest pressure 1~kPa (or $\kappa\simeq 181000$). 
Tests of the influence of viscous dissipation parameter $\zeta$ and maximum value $I_{\text{max}}$ of reduced strain rate in compression are 
also made for configurations compressed from a granular gas to 10~kPa.}
\label{tab:initP}
\centering
\begin{tabular}{|c|c|c|c|ccccccc|}  \cline{1-11}
Origin & P (kPa)& $\zeta$&$I_{\text{max}}$&$\Phi$ & $z^*$ & $x_0$ (\%) &  $x_2$ (\%) &$Z(2)$& $M_1$ &$M_2$\\
\hline
gas&$1$&$0.98$&$10^{-3}$&$0.5930\pm 0.0007$ &$4.546\pm 0.009$ & $11.1\pm 0.4$&$2.39$   &$1.58$&$0.160$&$0.217$\\
gas&$10$&$0.98$&$10^{-3}$&$0.5946\pm 0.0006$ &$4.59\pm 0.02$ & $10.2\pm 0.2$& $2.07$  &$1.59$&$0.159$&$0.213$\\
gas&$10$&$0.098$&$10^{-3}$&$0.5938\pm 0.0008$ &$4.61\pm 0.02$ & $10.9\pm 0.2$& $1.79$  &$1.57$&$0.150$&$0.194$\\
gas&$10$&$0.98$&$10^{-1}$&$0.5931\pm 0.0002$ &$4.60\pm 0.01$ & $10.2\pm 0.7$& $1.80$  &$1.59$&$0.159$&$0.212$\\
QS &$10$&$0.98$&$10^{-3}$&$0.5931\pm 0.0006$ &$4.641\pm 0.011$ & $10.1\pm 0.4$&$2.33$&$1.46$&$0.146$&$0.188$\\
gas&$100$&$0.98$&$10^{-3}$&$0.5975\pm 0.001$ &$4.69\pm 0.02$ & $8.9\pm 0.5$&$1.66$&$1.61$ &$0.153$&$0.197$\\
QS &$100$&$0.98$&$10^{-3}$&$0.5936\pm 0.0006$ &$4.79\pm 0.02$ & $8.6\pm 0.4$& $2.05$  &$1.40$&$0.138$&$0.178$\\
\hline
\end{tabular}
\normalsize
\end{table*}
As reported in table~\ref{tab:initP},
a value of the damping parameter ten times as small as the standard one $\zeta =0.98$ results in quite similar configuration properties, on compressing
a loose granular gas under $P=10$~kPa, with $I_{\text{max}}=10^{-3}$. So did in fact faster compressions, with $I_{\text{max}}=10^{-1}$ , keeping
$\zeta =0.98$.  The data of Table~\ref{tab:initP} thus suggest that we very nearly achieved the independence on dynamical parameters that is
expected in the $I\to 0$ limit with our choice of control parameters. 
We note, however, that other possible definitions of a \emph{random loose packing}, such as the one by 
Onoda and Liniger~\cite{OL90} result in different (smaller) solid fractions. 
Looser arrangements of equal-sized spherical particles can also be stabilized with adhesive contact forces, \emph{e.g.} on
introducing the capillary attractions produced by the menisci formed by a wetting fluid in the interstices between neighboring grains~\cite{KOH04}. 

In addition to packing fraction $\Phi$, coordination number $z^*$, fraction of rattlers $x_0$, 
Table~\ref{tab:initP} lists the reduced second moment $Z(2)$ of the normal
force distribution, as defined in ~\eqref{eqn:defza}, the proportion of two-coordinated beads 
(to be discussed in Section~\ref{sec:ten}), $x_2$, and the average values of
ratios $\vert \vert {\bf T}\vert \vert /N$ (friction mobilization)
among contacts carrying normal forces larger and smaller than the average, respectively denoted as $M_1$ and $M_2$. 
As a result of some
amount of quasistatic compression of the initial assembly, the width of the force distribution decreases, 
as witnessed by smaller values of $Z(2)$ in table~\ref{tab:initP}, and
so does the mobilization of friction, as measured by  $M_1$ and $M_2$. 
The effects of compression on the structure and the forces are further studied in paper II~\cite{iviso2}.
On comparing numerically simulated loose packings to experiments, it should be recalled 
that samples are assembled under low pressures in the laboratory: the hydrostatic
pressure under a 1~cm thick layer of glass beads is about $0.15$~kPa. 
Numerical configurations under higher confining pressures corresponding to mechanical tests in the laboratory
(\emph{e.g.}, sound propagation) are more appropriate models if the testing pressure is significantly larger than the initial, 
assembling pressure -- as for the ``QS'' samples of table~\ref{tab:initP}.

The effects of such proportions of rattlers in granular packings as reported in table~\ref{tab:initP} have to our knowledge never been studied in detail.
It should be emphasized that this relatively large population of rattlers does not jeopardize the global stability of equilibrium configurations,
as the stiffness matrix of the force-carrying network is found devoid of floppy modes (apart
from harmless, localized ones associated with two-coordinated particles, to be discussed in Section~\ref{sec:ten}). 

Our D samples should be compared with the simulations reported by Zhang and Makse~\cite{ZhMa05}, 
in which loose sphere packings were also prepared by isotropic compression. 
Those authors observed, in some cases, lower packing fractions than D values, $\Phi\simeq 0.57$. Their assembling method
is however different: they use a strain-controlled procedure, with a constant compression rate, and then relax the final state at constant volume.
In this approach, the pressure reaches very high levels, several orders of magnitude as large as the final value, 
before samples finally stabilize~\cite[Fig. 3]{ZhMa05}. Zhang and Makse report some dependence of the final state on the compressing rate.
Once translated into the dimensionless parameter $I$ we have been using here,
strain rates used in~\cite{ZhMa05}, defined with the typical pressure value $P=100$~kPa, range between $I=0.1$ and $I=100$. 
The slowest compression reported in~\cite{ZhMa05} is therefore
100 times as fast as the upper limit for $\dot \epsilon$ we have been enforcing in this work. Viscous forces also differ 
between the present simulations and those of Ref.~\cite{ZhMa05}, in which ``global damping'' terms are used (\emph{i.e.}, 
forces opposing the individual motion of particles, rather than relative motions). 
\subsubsection{Use of low friction coefficients: imperfect lubrication\label{sec:aaf0}}
One way to limit the effects of friction consist in lubricating the grains, 
as in the experimental study reported in~\cite{JM01}. 
If all intergranular friction could be suppressed in the assembling procedure, \emph{i.e.,} for perfect lubrication, 
then the structure of isotropic packings would be the one denoted as A, studied in Section~\ref{sec:asf}. 
The effect of a small friction coefficient in the contacts
while the grain assembly is being compressed can be regarded as a crude, simplified model for imperfect lubrication. 
We made samples, denoted as B, by compressing the granular gas, just like in the
A and D cases, with $\mu=0.02$. In order to approach the limit of slow compression rates better, we started from 
D configurations, decreased the friction coefficient to $\mu=0.02$, and then requested that $I<10^{-4}$ while
the samples got further compressed to equilibrium under 1~kPa ($\kappa\simeq 181000$). (In view of the results in the D case 
of Section~\ref{sec:aaf1}, we do not expect the final B state to be sensitive to damping parameter $\zeta$.)
We observed that this small friction 
coefficient had a notable effect on the final solid fraction, as the value 
$\Phi= 0.6270\pm 2.10^{-4}$ is significantly 
below the
frictionless (A) result, 
while the coordination number on the active structure is slightly reduced, 
down to $z^*= 5.75\pm 8.10^{-3}$, and the fraction of rattlers raised slightly, to $x_0 = (1.95\pm 0.02)\%$.
\subsubsection{Dense, frictional packings obtained by shaking\label{sec:vib}}
Another practical strategy to obtain dense configurations is to shake, vibrate or apply repeated ``taps'' on granular samples~\cite{NKBJN98,PB02,RPBBTB03}.
Such procedures involve the introduction of kinetic energy into already quite dense assemblies. In order to investigate their possible effects at a limited 
computational cost, we avoid the direct simulation of repeated shakes and adopted the following procedure. Starting from the dense A configurations (made without
friction and described in Section~\ref{sec:asf}), we first apply a homogeneous expansion, 
multiplying all coordinates by a common factor $\lambda$ slightly larger than $1$.
With equilibrated A states under confinement level $\kappa=39000$, the chosen value $\lambda = 1.005$ 
is more than enough to separate all pairs in contact. Then, in order to
mimic, in an idealized way, the motion set up by a shaking excitation, the beads are given random velocities (chosen according to a Maxwell distribution), and
interact in collisions which preserve kinetic energy, while the volume of the cell is kept constant. 
This ``mixing'' stage is simulated with the ``hard sphere molecular dynamics'' 
(event-driven) scheme (just like our initial granular fluids are prepared at $\Phi=0.45$, as described in Section~\ref{sec:MDres}). It is
pursued until each particle has had $n_{\text{coll}}=50$ collisions on average. The final preparation stage is a fast compression: velocities are
set to zero, particles regain their elastic and
dissipative properties (as defined in Section~\ref{sec:forces}, 
with friction coefficient $\mu=0.3$, and viscous dissipation, $\zeta=0.98$), the external pressure $P=10$~kPa
is applied \emph{via} the deformable periodic cell, until a final equilibrium is reached.

The final state is hereafter referred to as C. Quite unsurprisingly, its solid fraction, 
$\Phi=0.635\pm 0.002$, stays very close to the RCP value obtained in the A state.
However, the coordination number is considerably lower, $z^*=4.56\pm 0.03$, which is as small as the value obtained in the loose ($\Phi\simeq 0.593$) 
D state, while the proportion of rattlers
raises to $x_0=(13.3\pm 0.5)\%$. Remarkably, on comparing states B and C, the latter has a higher density, but a much lower coordination number, 
and a much higher fraction of rattlers. 

We did not thoroughly investigate the influence of parameters $\lambda$ and $n_{\text{coll}}$, introduced in the preparation procedure, on the resulting
C states. In the following we focus on configurations obtained with the values $\lambda=1.005$ and $n_{\text{coll}}=50$. Yet, we
noted that an increase of $\lambda$ to $1.01$ entailed only very slight changes of $\Phi$ and $z^*$ (which respectively decreased to $0.633$ and $4.54$), 
and that the suppression of the ``mixing'' stage (\emph{i.e.}, setting $n_{\text{coll}}$ to zero) resulted in much higher $z^*$ values (around $5.5$). 
Likewise, we did not check for a possible effect of $\zeta$ on the final state. 
Smaller values than the large one $\zeta=0.98$ used in our simulations of the
final compression stage of C sample
preparation are likely to have analogous effects to an increase of the duration of the agitated mixing stage, and should not increase the final coordination number.
\subsubsection{Global state variables: summary  and discussion}
\begin{table*}
\caption{Isotropic states ($\kappa\simeq 39000 $ for A and C, $\kappa\simeq 181000$ for B and D) for different assembling procedures.}
\centering
\begin{tabular}{|l|ccccccc|}  \cline{1-8}
Procedure & $\Phi$ & $z^*$ & $x_0$ (\%) &  $x_2$ (\%) &Z(2) & $M_1$ &$M_2$\\
\hline
\hline
A &  $0.6370\pm 0.0002$ &$6.074\pm 0.0015$ & $1.3\pm 0.2$&0&$1.53$&0&0\\
\hline
B ($\mu_0=0.02$) &$0.6271\pm 0.0002 $ &$5.80\pm  0.007$  & $1.95\pm 0.02 $&$\sim10^{-4}$&$1.52$&$0.016$&$0.018$\\
\hline
C ($\lambda=1.005$) &$0.635\pm 0.002 $ &$4.56\pm  0.03$  &$13.3 \pm 0.5$&$2.64$&$1.65$&$0.135$&$0.181$\\
\hline
D &$0.5923\pm 0.0006$ &$4.546\pm 0.009$ & $11.1\pm 0.4$&$2.39$&$1.58$&$0.160$&$0.217$\\
\hline
\end{tabular}
\label{tab:prep}
\normalsize
\end{table*}
Table~\ref{tab:prep} gathers some of the parameters characterizing
 the four different packing states studied in
the present paper. ($M_1$ and $M_2$ were defined in connection with table~\ref{tab:initP}). 
From table~\ref{tab:prep}, configurations with low coordination numbers (C and D) appear to exhibit a somewhat
wider normal force distribution (as measured by $Z(2)$), and a significant mobilization of friction,
with typical values of $\norm{\bf T}/N$ around $\mu/2=0.15$ for larger than average normal force components $N$ ($M_1$), 
and significantly above $\mu/2$ for smaller $N$ values ($M_2$).

The existence of C states shows that there is no systematic relationship between density 
and coordination number, contrarily to some statements in the literature~\cite{AsSaSe05}. Of
course, for one particular assembling method  both quantities will often vary in the same direction as functions of some control
parameter. For instance, on preparing samples by deposition under gravity, both density and 
coordination number are increasing functions of the height of free fall~\cite{ERCCD05}.
However, our results show that different preparation methods might lead to contrasting results. 

Our results about density and coordination number can be likened to observations made before in numerical models of sphere packings 
obtained with geometric construction rules~\cite{NOKA92}. The simplest versions of such
algorithms~\cite{VB72,JM87}, which mimic deposition under gravity, add particles one by one by dropping and rolling them
in contact with one or two previously deposited particles, until 
they are fixed when they rely on three contacts. Those produce
packings with $z=6$. More refined versions thereof~\cite{NOKA92,BaMe92} also involve other, more collective types of moves.
The final configurations then contain ``bridges'' or
``arches''~\cite{MeBaLu04}, defined as sets of particles the final stabilization of which is mutual and collective. 
In such arches, each grain relies on three others, but   
some pairs mutually rely, in part, on each other. Those ``bridged structures'' have much lower coordination numbers, down to about $4.5$. 

It is not clear, though, to what extent our configurations,
which were obtained within a full mechanical model, compare to those that result from such approaches. 
As shown, \emph{e.g.}, in~\cite{BRH02}, deposition algorithms based on
geometrical rules are supposed to ensure local stability properties, but the resulting granular pilings
might turn out to be globally unstable. Moreover, a description of our packings as a sequence of arches
placed one after another, assuming it is conceivable, would seem to contradict their homogeneity and isotropy: it is rather arbitrary, 
in isotropic packings, to regard some particles as ``relying'' on some others. Such a description was therefore not attempted. 
\subsection{Geometric characterization\label{sec:fgeom}}
\subsubsection{Pair correlation functions}
\begin{figure}[htb]
\includegraphics*[width=8.5cm]{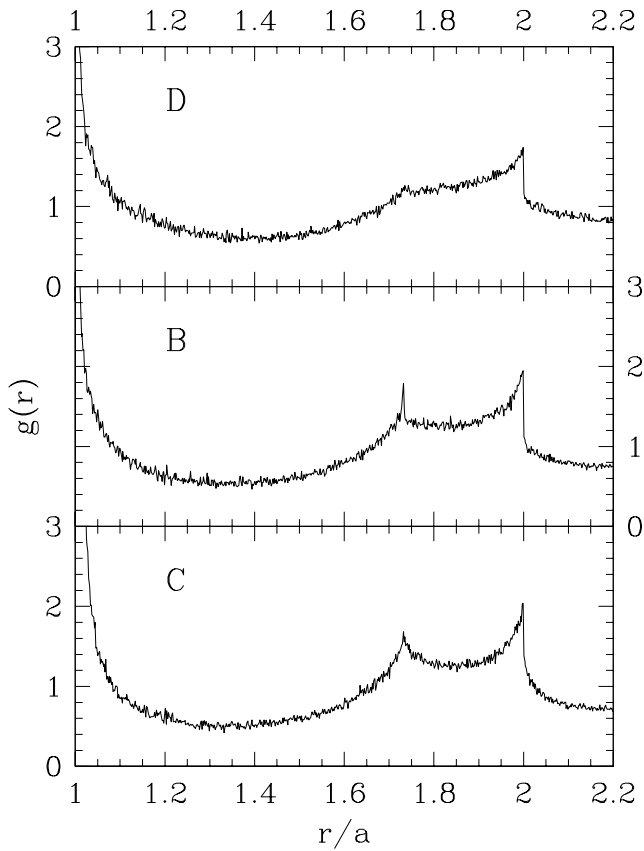}
\caption{\label{fig:grcomp}
Pair correlation functions $g(r)$ (definitions $g^{I}(r)$ and $g^{II}(r)$ coincide on this scale) for configurations B, C, D.
}
\end{figure}

As observed in previous experimental~\cite{AsSaSe05} and numerical~\cite{SEGHL02} results, pair correlation functions present the same features
at lower densities as at the largest one $\Phi\simeq 0.64$, in a weakened form, as shown on Fig.~\ref{fig:grcomp}.
On comparing those functions for states A to D, we observed what follows.
\bi
\im
C samples, obtained from A ones after small rearrangements, exhibit pair correlations that only differ 
in the detailed shape of the peaks (\emph{e.g.} below $1.05a$), and 
is indistinguishable elsewhere.
\im
In spite of the large number of rattlers in samples C and D, $g^{I}(r)$ and $g^{II}(r)$ (as defined in Section~\ref{sec:asfbb})
cannot be distinguished on the scale of Fig.~\ref{fig:grcomp}.
\im
The depth of the trough between $r/a = 1.1$ and $r/a = 1.5$ increases with density.
\im
The integral below the peaks correlates with density, but the height and sharpness of the \emph{drop} at $r/a=\sqrt{3}$ and $r/a=2$ correlate with coordination
number (which is larger for B than for C), in agreement with the interpretation of such features suggested in Sec.~\ref{sec:asfgeom}. 
\ei
\subsubsection{Near neighbor coordination numbers}
\begin{figure}[htb]
\includegraphics*[angle=270,width=8.5cm]{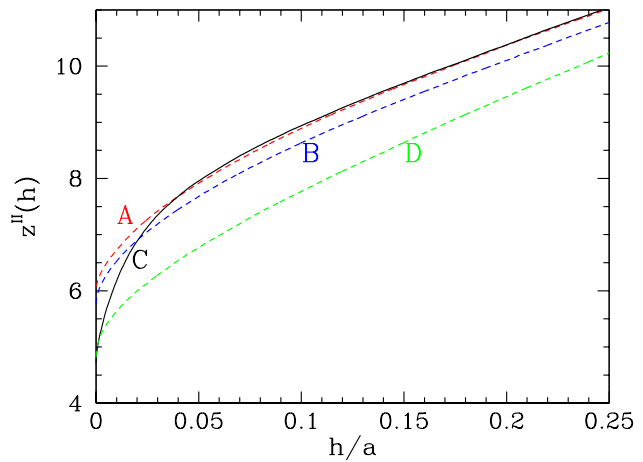}
\caption{\label{fig:zhlinABCD}
(Color online) Coordination number for neighbors at distance $\le h$, $z^{II}(h)$, for configurations A (red, upper dashed line), B (blue, middle dashed line),  
C (black, solid line) and D (green, bottom dashed line).
}
\end{figure}

The gap-dependent coordination number $z^{II}(h)$ is shown on Fig.~\ref{fig:zhlinABCD} for samples A to D. 
Fig.~\ref{fig:zhlinABCD} shows that, as might have been intuitively expected, 
$z(h)$ correlates with coordination number for small $h$ and with density for larger distances, $h\ge 0.04 a$.
We  preferably use definition $z^{II}$, which is obtained on 
bringing the rattlers in contact with the backbone with small, random forces, as explained in Sec.~\ref{sec:asfbb}.  $z^{II}(h)$
can be thought of as more physically meaningful than $z^I(h)$, which directly results from the simulation of the packing, and is 
somewhat ambiguously defined because the positions of the rattlers are not specified. 
Functions $z^{II}(h)$ corresponding to B and C states cross each other for $h\simeq 0.02 a$.

$z^{I}(h)$ and  $z^{II}(h)$ might be fitted by power laws: 
\be
\ba
z^{I}(h) &= B^{I} + A_{I}h^{\beta_{I}}\\
z^{II}(h) &= B^{II} + A_{II}h^{\beta_{II}}\\
\ea
\label{eqn:zIIh}
\ee

\begin{figure}[htb]
\includegraphics*[angle=270,width=8.5cm]{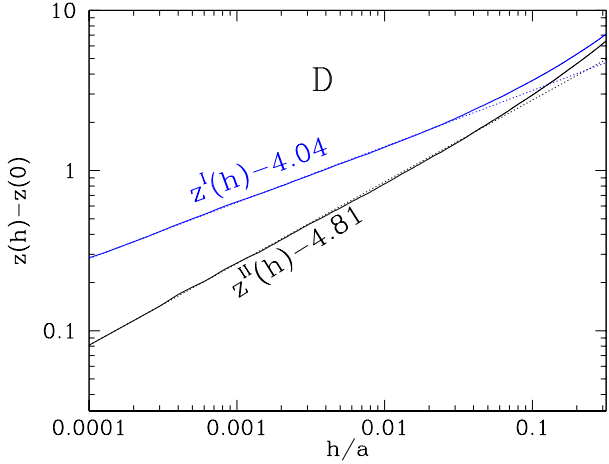}
\caption{(Color online) $z^I(h)-z$ (black) and $z^{II}(h)-z^{II}(0)$ (blue) versus $h$ on doubly logarithmic plot for D samples at lowest pressure. 
The slopes of the corresponding dotted straight lines (power law fits) are $\beta_{II}=0.51$ and $\beta_{I}=0.35$ (see Eqn.~\ref{eqn:zIIh}).
\label{fig:zhD}}
\end{figure}

\begin{figure}[htb]
\includegraphics*[angle=270,width=8.5cm]{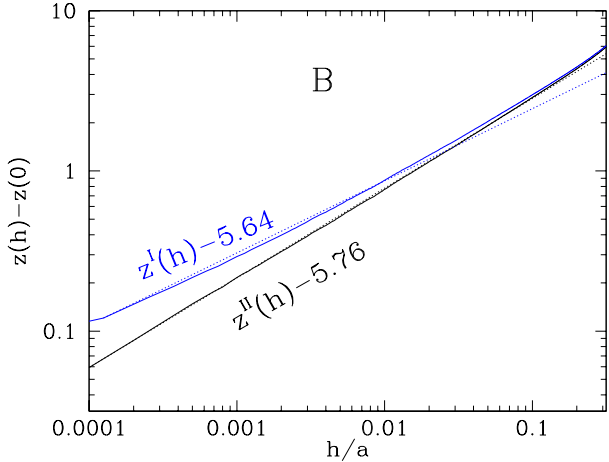}
\caption{(Color online) Same as Fig.~\ref{fig:zhD}, for B samples. 
Dotted lines have slopes $\beta_{II}=0.56$ and $\beta_{I}=0.45$.
\label{fig:zhB}}
\end{figure}

\begin{figure}[htb]
\includegraphics*[angle=270,width=8.5cm]{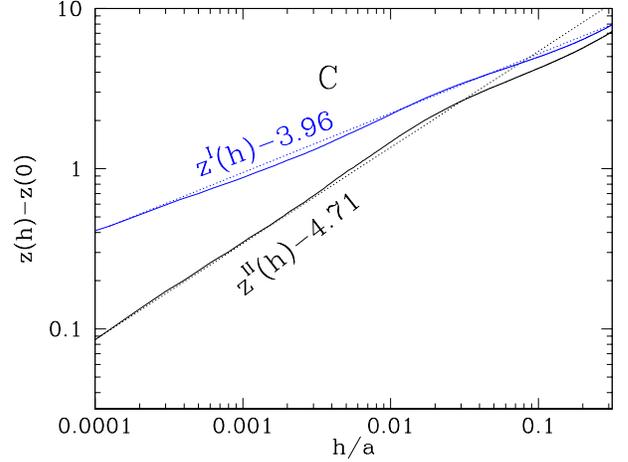}
\caption{(Color online) Same as Fig.~\ref{fig:zhD}, for C samples. 
Slopes of dotted straight lines: $\beta_{II}=0.6$ and $\beta_{I}=0.37$.
\label{fig:zhC}}
\end{figure}

Figs~\ref{fig:zhD},  \ref{fig:zhB} and \ref{fig:zhC} display $z^I(h)-B^{I}$ and $z^{II}(h)-B^{II}$ as functions of $h$ 
on logarithmic plots for samples D, B and C (due to the influence, at short distance, of contact deflections on $z(h)$ data, fit parameters
$B^{I}$ and $B^{II}$ are a little smaller than $z$ and $z^{II}(0)$).  For $h$ values smaller than $10^{-4}$, the
lowest limit on the axis on Figs.~\ref{fig:zhD}, \ref{fig:zhC} and \ref{fig:zhB}, the gap $h$ is of the same order as elastic deflections, $\kappa^{-1}$, 
and we do not observe simple power laws (as on Fig.~\ref{fig:zh}).
These figures (on which the corresponding values of $z\simeq z^I(0)$ and $z^{II}(0)$ are also provided) show that $z^I$ and $z^{II}$ have quite different $h$
dependences. This should be accounted for on studying the closing of contacts due to compression (see paper II~\cite{iviso2}). As a result
of the computation leading to the equilibrated state, many pairs of neighbors end up separated by a very small interstice, so that $z^I-z$ already reaches values
larger than $0.3$ for $h=10^{-4}a$ in samples C and D. $z^I$ then grows with $h$ more slowly than $z^{II}$, 
with $\beta_{II}>\beta_{I}$, and a power law fit of lesser quality. Exponent $\beta_{II}$, 
which should be regarded as a more intrinsic quantity than $\beta_I$, appears to correlate
with solid fraction. It has the same value $0.6$ in samples C and A (Figs~\ref{fig:zhC} and \ref{fig:zh}), 
decreases to $0.55$ in the intermediate density state B, and to $0.51$
for the least dense one, D. The power-law form of $z^{II}$ extends to $h\simeq 0.3$ in configuration A, 
to about  $h\simeq 0.2$ in configuration B, and only to $0.04$ and $0.05$ 
for C and D. Such limited power law ranges preclude comparisons with the experimental data of~\cite{AsSaSe05}.

One has $z^{I}(0)\simeq z$ as a very good approximation, since contacts carrying no force in the equilibrated state obtained by MD are very scarce. 
$z^{II}(0)$, on the other hand, is the \emph{geometric} coordination number once all rattlers are pushed against the backbone. It is larger than the 
\emph{mechanical} coordination number, $z$. Specifically, because all rattlers, in treatment II,
are dealt with as frictionless, one has:
\be
z^{II}(0) = z + 6x_0 - \frac{2N_{rr}}{n},
\label{eqn:relzIIz}
\ee
in which $N_{rr}$ is the final number of contacts between rattlers once they are positioned against the backbone. Value of $z^{II}(0)$ can be read 
on Fig.~\ref{fig:zhlinABCD}. As pointed out in Section~\ref{sec:ratt}, they can be compared to coordination number values of samples under gravity.
The results of Silbert \emph{et al.}~\cite[Figs. 2 and 3]{SEGHL02}, with $\mu=0.3$, 
correspond approximately with our D samples: $\Phi\simeq 0.59$ and $z^{II}(0)\simeq 5$.
The systems simulated in~\cite{SEGHL02} are however not isotropic. None of the samples made under gravity in~\cite{SEGHL02,SGL02}
appears to resemble our C state.
\subsubsection{Absence of crystalline order, local order parameters}
All indicators of incipient crystalline order given in Table~\ref{tab:aacryst} for frictionless A samples take lower values in states B and D, while
C configurations, due to their geometric proximity, are close to A ones in this respect. Already scant in dense configurations assembled without friction, 
traces of crystallization are thus negligible in looser ones obtained with frictional beads. Like for A samples, numerical data on configurations around
one sphere $i$, as characterized  by the pair $(\hat Q_4(i),\hat Q_6(i))$, used by Aste \emph{et al.}~\cite{AsSaSe05}, 
are presented for states B, C, and D and compared to their experimental results in Appendix~\ref{sec:appaste}. It is observed that local disordered environments around 
one grain are very similar in
numerical and experimental configurations of equal densities, while local HCP-like arrangements occur 
with similar (low) frequencies, and FCC-like ones are present in the laboratory,
but not detected in simulations. 
\subsection{Properties of force networks\label{sec:ten}}
\subsubsection{Local contact coordination numbers}
The distribution of local coordinations is given in Table~\ref{tab:coordBCD},
\begin{table}[htb!]
\centering
\begin{tabular}{|c|ccccccccccc|}  \cline{1-12}
State& $x_0$ & $x_1$& $x_2$& $x_3$& $x_4$& $x_5$& $x_6$& $x_7$& $x_8$& $x_9$& $x_{10}$\\
\hline
\hline
B (I) & $1.95$ & $0$& $0.05$& $0.5$& $16.3$& $26.9$& $27.5$& $18.4$& $7.1$& $1.4$& $0.1$\\
\hline
C (I)& $13.3$ & $0$& $2.6$& $15.1$& $26.5$& $23.4$& $13.2$& $4.8$& $0.9$& $0.15$& $0$\\
\hline
D (I)& $11.1$ & $0$& $2.4$& $13.8$& $29.1$& $25.6$& $13.3$& $4.0$& $0.7$& $0.03$& $0$\\
\hline
B (II) & $0$ & $0$& $0$& $2.1$& $15.3$& $26.5$& $27.5$& $18.9$& $7.8$& $1.6$& $0.2$\\
\hline
C (II)& $0$ & $0$& $0$& $18.9$& $22.8$& $27.0$& $18.6$& $9.3$& $2.8$& $0.5$& $0.1$\\
\hline
D (II)& $0$ & $0$& $0$& $17.2$& $25.4$& $29.0$& $18.5$& $7.7$& $1.9$& $0.3$& $0$\\
\hline
\end{tabular}
\normalsize
\caption{Percentage $x_i$ of grains having $i$ contacts in configurations B, C and D, on ignoring contacts with or between
rattlers (I), or on fixing them onto the backbone with small, random forces (II).
\label{tab:coordBCD}}
\end{table}
for both mechanical (I) and geometric (II) definitions of contacts, for states B to D. Compared to the A case (Table~\ref{tab:coord}), the distribution is
 shifted to lower values, with 4 and 5 the most frequent ones (rather than 6) in low coordinated packings C and D. Those samples also have quite a large population of
three-coordinated spheres, and a notable one of divalent (two-coordinated) particles. This contrasts with the frictionless case for which $x_2=x_3=0$. Without friction,
divalent spheres, from Eqn.~\ref{eqn:relhk0}, written with $N_c=2$, $N_f=3$, $h=0$, would 
imply a mechanism and therefore an instability, and three-coordinated ones, in the absence of external forces and cohesion, cannot
be equilibrated by non-vanishing normal forces the net effect of which necessarily 
pushes them away from the plane defined by the three centers of their touching neighbors (the
non-generic case with the four sphere centers within the same plane leading to an instability). 
Spheres with three contacts therefore need some mobilization of friction to transmit non-vanishing forces in an equilibrium configuration, for tangential components
are requested to cancel this net repulsion. The Coulomb condition then restricts such possible configurations to flat enough tetrahedra for contact forces to remain
within the friction cone. This explains the small value of $x_3$ in low friction ($\mu=0.02$) B samples. 
\subsubsection{The special case of divalent grains~\label{sec:dival}}
\begin{figure}[htb]
\includegraphics*[width=8.5cm]{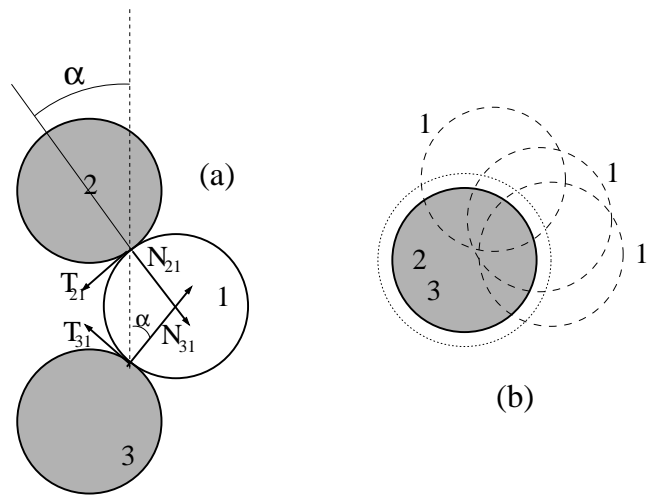}
\caption{\label{fig:dival}
Equilibrium and free motion (mechanism) of one sphere (marked 1) with two contacts (with particles marked 2 and 3). 
(a): in the plane of the three centers, normal and
tangential components of the two contact forces balance along the line joining the contact points (dotted line). 
(b): seen from above, along the direction of the line
of the centers of spheres 2 and 3, sphere 1 can move and occupy the different positions depicted with dashed lines, 
its center describing the dotted circle around the 2-3 axis.}
\end{figure}

With friction, the small structure formed by one sphere having two contacts with fixed objects (Fig.~\ref{fig:dival}), due to Eqn.~\ref{eqn:relhk}, 
in which the number of degres of freedom (6) is equal to the number of contact force coordinates, has a degree of force indeterminacy equal to its number of independent 
mechanisms: $\hhh =k$. In fact, both numbers are equal to 1. Self-balanced contact forces (see first part of Fig.~\ref{fig:dival}) 
are oriented along the line joining the two contacts,
just like in the corresponding 2D case dealt with in~\cite{McGaHe05}, and their amplitude is a free parameter (the degree of
``wedging'' of the grain in the corner formed by its two neighbors~\cite{McGaHe05}). Such a possibility requires in practice that the angle, which we denote as 
$\alpha$  (Fig.~\ref{fig:dival}), between the line
joining the centers of 1 and 2 and the one joining the contact points be smaller than the angle of friction, for the total contact force to stay within the Coulomb cone.
In C and D samples, we observed $\tan \alpha$ to be distributed rather evenly between $0$ and $\mu$, 
while the intensity of forces transmitted by divalent spheres ranged from $0$ to a few times $Pa^2$. 
The mechanism associated with divalent spheres is a free rolling motion on the two contacts, 
the line joining the contact points being the instantaneous axis of rotation, 
as shown on fig.~\ref{fig:dival}(b). In this motion, it is readily checked (see Appendix~\ref{sec:appdival}) that the rules given 
in Appendix~\ref{sec:appendixTrot} for the evolution of contact forces in the case of
rolling and pivoting specify that contact forces will remain carried by the line joining the two contacts, 
with a constant intensity, as such contacts move on the
surface of the fixed spheres. Such forces will
do no work and the kinetic energy of the mobile sphere as well as the elastic energy stored in its two contacts will be kept constant. 
The equilibrium of the divalent particle
is thus marginally stable, with matrix $\stib$ causing zero acceleration to the free motion. 
However, such a motion does affect the balance of \emph{moments} on spheres
marked 2 and 3 on Fig.~\ref{fig:dival}, since the constant force is applied at a point that is moving on their surface. 
Therefore the stability of such free motions, as regards the
global contact network, requires some additional analysis -- which is tackled in Appendix~\ref{sec:appdival}, 
where it is concluded that the packing remains stable. Unlike in
frictionless packings~\cite{SA98,JNR2000}, mechanisms in the presence of friction do not necessarily lead to instabilities. 
On building the constitutive stiffness matrix $\stia $
in the samples we studied, we could check that no other mechanism was present on the backbone than those rolling motions of 
divalent spheres. Once some stiffness element is introduced to impede the free motion of divalent spheres, one can \emph{e.g.} check that 
the Cholesky factorization of the stiffness matrix only involves strictly positive terms on the diagonal. 

There can be no contact between two divalent spheres, as the simultaneous equilibrium of each of them
with two forces carried by the line joining its two contact points, as on Fig.~\ref{fig:dival}, is impossible. 
\subsubsection{Degree of force indeterminacy}
On the backbone, with $n(1-x_0)$ spheres and $N_f^*=6n(1-x_0)+3$ degrees of freedom, 
one has on average $z^*n(1-x_0)/2$ contacts and $k=3+x_2 n$ independent mechanisms, hence a degree
of force indeterminacy (hyperstaticity), from~\eqref{eqn:relhk}, given by:
\be
\ba
\hhh&=N_f^*\frac{z^*-z^* _0}{4}=N_f^*\frac{z^{**}-4}{4}, \ \text{with}\\
z^* _0&=4-\frac{2x_2}{3(1-x_0)}\ \text{and}\\
z^{**} &= z^*+\frac{2x_2}{3(1-x_0)}
\label{eqn:hyperf}
\ea
\ee
The backbone is devoid of force indeterminacy ($\hhh=0$) when its coordination number is equal to
$z^* _0$.
Because of the mechanisms associated with divalent beads $z^* _0$ is strictly smaller than 4. Alternatively, one can define
a ``corrected'' backbone coordination number $z^{**}$, as written in~\eqref{eqn:hyperf}, which is equal to 4 in
the absence of force indeterminacy. 
As to the global mechanical coordination number $z_0$ corresponding to the absence of force indeterminacy, its value, given by
\be
z_0=4(1-x_0)-\frac{2x_2}{3},\label{eqn:zminf}
\ee
is well below 4. From the data of table~\ref{tab:coordBCD}, $z_0$ is about $3.45$ in state C and $3.54$ for D
(while $z$ is close to 4). Although $\hhh$ is relatively small compared to the number of degrees of freedom $N_f$,
the samples with friction and low coordination are still notably hyperstatic -- a conclusion shared by other studies~\cite{SEGHL02}, 
which we reach here in the slightly different context of packings in a uniform state of stress. Unlike for frictionless sphere assemblies, there is actually no
special reason to expect packings with intergranular friction to become isostatic in the rigid contact limit. The essential difference is that contact forces
can no longer be regarded as enforcing hard geometric constraints like impenetrability, and hyperstatic configurations do not require exceptional arrangements or
matching of particle sizes as in the frictionless case~\cite{OR97a,MO98a,TW99,JNR2000}. 

Unlike us, Zhang and Makse~\cite{ZhMa05} speculate that isostatic packings could be obtained in the limit of slow compressions of samples with ordinary values of
$\mu$.  This is however due to a divergence of 
interpretation, rather than a contradiction in numerical results, 
since their minimum coordination numbers $z^*$, excluding rattlers, are similar to ours, $z^*\simeq 4.5$. 
Zhang and Makse could only approach 
configurations devoid of hyperstaticity on setting the friction coefficient to infinity (see Section~\ref{sec:muinf} below). 
The degree of force indeterminacy per degree of freedom on the backbone is still equal, from~\eqref{eqn:hyperf}, 
to $0.141$ in D samples and $0.145$ in C ones, and varies very little with compression rate in the range we explored, which
extends to significantly smaller values than the ones used in~\cite{ZhMa05}, as stressed above. It is not obvious whether special 
experimental situations might occur in which real granular assemblies approach vanishing degrees of hyperstaticity.
\subsubsection{Distribution of normal forces}
Since the force-carrying structure maintains a non-vanishing degree of force indeterminacy even in the rigid limit for
frictional packings, the force distribution in states B, C, and D, unlike in A configurations, is no longer
a geometrically determined quantity in the rigid limit.
 
\begin{figure}[htb]
\includegraphics*[angle=270,width=8.5cm]{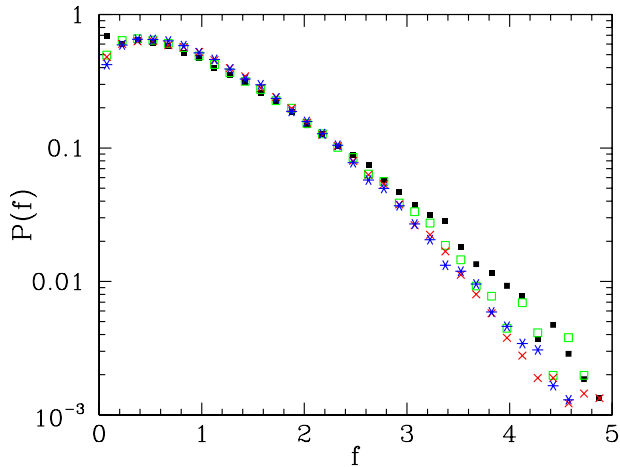}
\caption{\label{fig:histout}
(Color online) Distribution of normal forces, normalized by its average, $f=\frac{N}{\ave{N}}$, in states A (red crosses), 
B (blue asterisks), C (black square dots) 
and D (green open squares) at the lowest simulated pressure.
}
\end{figure}

The distribution of normal components of contact forces (normalized by its average $\ave{N}$) is shown on 
Fig.~\ref{fig:histout} for all 4 configurations A, B, C, and D, at the lowest 
pressure (as given in Table~\ref{tab:prep}), at the end of the assembling process.
We observe, as in many other numerical~\cite{RJMR96,OR97b,Antony00,SGL02} and
experimental~\cite{MJN98,BMMJN01} studies, an approximately exponential decay of $P(f)$ for large values, 
which is somewhat slower in states with low coordination number
(in agreement with the values of $Z(2)$ given in Table~\ref{tab:prep}). 
It should be pointed out, however, that much larger differences between force
distributions in the four studied states A, B, C, D will appear on increasing the confining pressure (see~\cite{iviso2}, paper II). 
This is already apparent in the dependence
of $Z(2)$ on the previous history of D samples in table~\ref{tab:initP}. 

All probability distribution functions show a local minimum for $f\to 0$, except in state C. 
Although it was remarked in past publications~\cite{RWRJM97} that an upturn of the p.d.f. 
at low forces, as for C configurations, appeared when packings were not fully equilibrated, 
C configurations satisfy equilibrium requirements as well as the others.
\subsubsection{Friction mobilization}
As frictionless sphere packings are unstable for $z^*<6$~\cite{JNR2000}, 
a certain level of friction mobilization is required in states B to D, even under isotropic stresses. In
particular, non-vanishing
tangential forces are indispensable to ensure the equilibrium of grains with 2 and 3 contacts, 
for which they relate to the local geometric configuration, as discussed above
(see \emph{e.g.,} Fig.~\ref{fig:dival}). Some information about friction mobilization is given in Table~\ref{tab:mob23}.
\begin{table*}[htb!]
\centering
\begin{tabular}{|c|ccccccccccc|}  \cline{1-12}
State&$X_1$&$M_1$&$M_2$&$N^{(2)}$&$X_1^{(2)}$&$M_1^{(2)}$&$M_2^{(2)}$&$N^{(3)}$&$X_1^{(3)}$&$M_1^{(3)}$&$M_2^{(3)}$\\
\hline
B&$0.42$&$0.016$&$0.018$&n. d.&n. d.&n. d.&n. d.&$0.23$&$0.09$&$0.02$&$0.01$\\
C&$0.41$&$0.14$&$0.18$&$0.58$&$0.19$&$0.14$&$0.17$&$0.24$&$0.09$&$0.13$&$0.04$\\
D&$0.41$&$0.16$&$0.22$&$0.71$&$0.22$&$0.16$&$0.18$&$0.26$&$0.10$&$0.15$&$0.05$\\
\hline
\end{tabular}
\normalsize
\caption{
For states B, C, and D, at the lowest pressure (see table~\ref{tab:prep}), proportion $X_1$ 
of contacts carrying normal forces larger than the average $\ave{N}$, average
ratio $\frac{\norm{ {\bf T}}}{N}$ for contacts with $N> \ave{N}$ (respectively, $N<\ave{N}$) $M_1$ (resp. $M_2$). 
The same quantities with superscripts $(2)$ and $(3)$
apply to contacts implying spheres of coordination 2 and 3. $N^{(2)}$, $N^{(3)}$ 
are the average normal forces carried by these two populations of contacts, normalized by 
$Pa^2$.  Note that divalent
grains are absent in state B, and the corresponding quantities are therefore not defined.
\label{tab:mob23}}
\end{table*}
In Table~\ref{tab:mob23} we distinguish between contacts carrying normal forces larger and smaller 
than the average.
Those two populations of contacts, as first distinguished in~\cite{RWRJM97,RWJM98}, as the ``strong'' and ``weak'' networks, are
attributed different roles, especially under anisotropic stresses. We merely use these categories
here to gather information, in a compact, summarized form, about some aspects of force
networks, which correlate to the force level.
Table~\ref{tab:mob23} shows that, although three-coordinated particles tend to carry small forces on average, 
a notable proportion of them, and of the divalent ones, participate in the
strong ``force chains'' with larger than average force levels. Friction mobilization is necessary in the contacts with such spheres,
and reaches similar levels on average in the whole network. It it larger for contacts carrying small loads.
\subsubsection{The limit of large friction coefficients\label{sec:muinf}}
Motivated by the search for an isostatic limit in packings with intergranular friction, 
Zhang and Makse~\cite{ZhMa05} assembled numerical packings with friction coefficients
equal to infinity. Then they could get $z^*\simeq 4.15$. In order to investigate, in paper III~\cite{iviso3}
the elastic properties of tenuous contact networks, we also prepared a set of 5 configurations 
on compressing a granular gas
under $P=1$~kPa with $\mu= +\infty$ and condition $I\le 0.001$. These states,
herafter referred to as Z configurations, have
solid fraction $\Phi=0.5917 \pm 0.0008$, backbone coordination number $z^*=4.068 \pm 0.006$, proportion of rattlers $x_0 = (18.4\pm 0.5)\% $ 
and of divalent beads $x_2 = (6.8\pm 0.5)\%$, and $\hhh$ is indeed small in that case, $\frac{\hhh}{N_f^*}\simeq 0.031$. 
It seems therefore very plausible that the degree of hyperstaticity
vanishes in this case, as $\kappa \to\infty$, for slowly assembled packs. In the light of this observation, the absence of such a limit
for finite $\mu$ can be attributed to sliding friction destabilizing barely rigid structures, which collapse and tend to form contacts in excess over the
minimum count.
\subsection{\label{sec:aafdisc} Conclusions}
We summarize here the most salient results of Section~\ref{sec:assemblef}, about systems with friction.
\subsubsection{Diversity of states of equilibrated packings}
The variety of inner structures of isotropic bead packings we have obtained, as summed up in Table~\ref{tab:prep}, 
shows that the solid fraction is not the only variable determining the internal state of an isotropic equilibrated packing. 
In particular, the backbone coordination number $z^*$ can vary throughout the whole interval from about 4.5 to 6 in  
the rigid limit ($\kappa\to+\infty$), in packings with a solid fraction close to the RCP value. Some systems can be denser than others, but with 
a considerably smaller coordination number. Systems compacted by vibration 
should have smaller coordination numbers than systems assembled with low friction coefficients.
\subsubsection{Geometry and length scales\label{sec:length}}
Geometric characteristics corresponding to length scales above about 4 or 5\% of 
the particle diameter correlate with density: this applies to the global shape and the area
under the peaks of pair correlation function $g(r)$, to near neighbor gap-dependent 
coordination number $z(h)$, and to local neighbor arrangements around one bead, as measured
by the local order parameters charted in Appendix~\ref{sec:appaste}. 

Features associated with smaller scales, such as $z(h)$ for $h\le 0.04 a$ or the shapes of the peaks of $g(r)$ -- 
part of which approach a discontinuity in the rigid limit -- correlate
with coordination number.

Finally, a third, smaller length scale $\kappa^{-1}a$ is associated with contact deflection and vanishes in the rigid limit. 
On this scale the geometric properties of the packing depend on the contact law.
\subsubsection{Role of rattlers}
Rattlers represent a significant fraction (above 10\%) of the particles in poorly coordinated systems, although the packings are stable. 
The treatment of rattlers -- 
whether they are left floating in arbitrary positions (I)
or gently pushed against the backbone by small forces (II) -- 
changes geometric data on the intermediate scale mentioned in the previous paragraph (Sec.~\ref{sec:length}), 
such as the exponent of a power-law fit of $z(h)$. Treatment II should
be preferred if comparisons are to be made with packings under gravity. 
It leads to the definition of a ``geometric'' coordination number, $z^{II}(0)$, different from the
mechanical one ($z$).
\subsubsection{Influence of micromechanical parameters}
Our data suggest (table~\ref{tab:initP}) that the states obtained on isotropically compressing a granular 
gas no longer depend on viscous damping parameter $\zeta$ in the limit of 
slow compression ($I\to 0$). If the true value of the friction coefficient is used at this stage (\emph{i.e.}, without ``lubrication''), 
the resulting state (our D configurations)
can be regarded as a reference, loose packing limit of this assembling method. Other methods nevertheless result in lower densities.
\subsubsection{Force indeterminacy}
In samples assembled by isotropic compression, the backbone does not lose its force indeterminacy 
at low pressure, even in the slow compression limit, except for $\mu\to+\infty$.
The degree of force indeterminacy, $\hhh$, decreases to about 14\% of the number of backbone degrees of freedom in poorly coordinated systems with $\mu=0.3$
On computing $\hhh$ it is necessary to take into account the contribution of divalent grains, which define localized (harmless) floppy modes.
Consequently the value of the backbone coordination number $z^*$ corresponding to $\hhh=0$ is slightly smaller than 4.
\section{Discussion and perspectives\label{sec:concl}}
The most important novel feature of our simulation results is the wide variability of coordination numbers for the same density in isotropic packings.
Most often, dense numerical samples are obtained on suppressing friction: A-type packings are simulated, in which the coordination number is high. 
C-type systems have about the same density, but a coordination number as low as in the loosest states D obtained by direct compression. In order 
to simulate dense laboratory samples, should one use A or C configurations ? The answer depends of course on how laboratory samples are assembled.
Our results show that vibrated ones are likely to have smaller coordination numbers than lubricated ones for the same density. In paper III, 
we compare elastic properties of our B and C states to the ones measured by Jia and Mills~\cite{JM01} on glass bead samples either
compacted by shaking the container or assembled with a lubricant.

In the X-ray tomography experiments by Aste~\emph{et al.}~\cite{ASSS04,AsSaSe05}, 
sphere packings are imaged with a 
resolution (voxel size) of about 4\% of nominal diameter $a$, while the 
diameter distribution extends at least to $\pm 0.03 a$. 
In spite of serious efforts to eliminate the influence of size distribution by deconvoluting correlation data, their
estimation of coordination numbers are well above the upper limit 6 in dense samples, which is in principle impossible under a low pressure. 
It seems that such experiments only provide access to the largest of the three length scales mentioned in Section~\ref{sec:length}, and are thus
unable to distinguish between A-like and C-like microstructures. 
We shall see in paper III~\cite{iviso3} that measurements of elastic
moduli are much better suited to obtain information on coordination numbers by experimental means. 

However, it is first necessary to assess the
influence of the pressure level on the packing inner states. As hinted by the results 
of Table~\ref{tab:initP}, a quasistatic compression affects the force distribution
and the level of friction mobilization. Elastic properties being usually measured above a certain confinement level (typically, a few tens
of kPa), the necessary study of the effects of a quasistatic compression is carried out in paper II~\cite{iviso2}.  

Beyond the elastic properties, which characterize the response to small load increments, 
the quasistatic, elastoplastic mechanical behavior of packings prepared with different
microstructures should also be studied. 
Peak deviator strength and dilatancy normally correlate with initial packing density~\cite{DMWood,BiHi93,MIT93,TH00,RR04}. 
But for one given density, what is the influence of the coordination number on the stress-strain curves~?

Granular systems are often packed under gravity by pouring samples in containers, and such processes, which do not necessarily 
produce homogeneous states~\cite{RT87} should be studied by numerical simulations too, and the analysis of
dynamical effects in the assembling stage should be pursued. Adhesive 
contact forces, as in
wet granular assemblies~\cite{KOH04}, can also greatly affect the preparation of solid granular packings~\cite{GiRoCa07}. 
Other perspectives to the present work are the investigation of the microstructure of polydisperse systems, 
and of assemblies of non-spherical particles~\cite{Donev-ellipse}. 
\appendix
\section{Contact elasticity and friction \label{sec:appendixfnft}}
The contact law between spherical elastic bodies, with a Coulomb criterion for friction applied locally (to
the surface force densities), leads to complicated history-dependent force-displacement
relationships~\cite{JO85,MiDe53}. Even in some cases with no slip anywhere in the contact region, the tangential
stiffness $K_T$ of a contact was shown~\cite{JoNo97} to depend on the past history of the contact loading, and
to change according to the direction of the displacement increment. Strictly speaking, the response of intergranular
contacts, even to arbitrary small load increments, should not be termed ``elastic''. The simplified law we adopted 
involves a tangential stiffness $K_T$ depending on the normal deflection $h$, but independent of the current
mobilization of friction. This is the same
approximation as used in~\cite{MGJS99,Makse04}: the value of $K_T$ is the correct one in the absence of
elastic relative tangential displacement, when ${\bf T}=0$. 

However, as stressed in~\cite{EB96}, such a model is thermodynamically inconsistent, for the elastic energy might increase
at no cost. Consider, \emph{e.g.,} quasistatically reducing $h$ at constant $\delta {\bf u}_T$, 
thereby, according to this contact model, reducing normal force $N$ at constant ${\bf T}$, without reaching
the Coulomb limit. The recoverable elastic
energy stored in the contact is given by
\be
w = \frac{2}{5} \tilde E \sqrt{a} h^{5/2}+ \frac{1}{2}\frac{{\bf T}^2}{K_T},
\label{eqn:defw}
\ee
which grows as $K_T$ decreases, without the external force doing any work, thus implying a net creation of energy.
To avoid such effects, ${\bf T}$ is rescaled (as advocated by O. Walton~\cite{Walton}), 
whenever $N$ decreases to $N-\Delta N$, down to  ${\displaystyle {\bf T} \frac{K_T(N-\Delta N)}{K_T(N)}}$, before accounting for
tangential relative displacement increments. No such rule applies to increasing normal force cases. Such a procedure
was shown by Elata and Berryman~\cite{EB96} to systematically
produce energy dissipation in cyclic loadings of the contact. 

Such peculiarities of the contact law affect the form and, in fact, the very definition of an elastic response of the contact network, 
an issue which will be discussed in paper III.

\section{Transport of contact forces due to particle motion \label{sec:appendixTrot}}
In molecular dynamics calculations, as well as in static approaches (as outlined in Section~\ref{sec:rist})
one has to relate small contact force increments in any contact to
 the small displacements ${\bf u}_i$ and rotations $\Delta {\bf \theta}_i$ of the grains. 

Increments 
$\Delta (N_{ij} \bf{n}_{ij})$ and $\Delta \Tij $ of the normal
and tangential parts of the force in the contact between grains $i$ and $j$
have two different origins: they stem from the contact law, as written down in Section~\ref{sec:model}, and also
from the motion of the particle pair. As the grains move, so does the deformed contact region,
and therefore the resulting contact force changes. The relevant formulae are derived and written below for
small increments in the static case. For dynamical
computations, displacements are to be replaced by velocities, and increments by time derivatives.

The normal force variation is simply
\be
\Delta (N_{ij} \nij ) = \Delta N_{ij}\nij + N_{ij} \Delta \nij ,
\label{eqn:deltafn}
\ee
with 
\be
\Delta \nij  = \frac{1}{\vert\vert {\bf r}_{ij}\vert\vert }\left(\ww{1}-\nij \otimes
\nij \right)\cdot\left(\tilde {\bf u}_j -\tilde {\bf u}_i -\ww{\epsilon}\cdot {\bf r}_{ij}\right),
\label{eqn:deltan}
\ee
while $\Delta N_{ij} $ is related by the Hertz law to the variation in the normal deflection of the 
contact.

For the tangential reaction we introduce the decomposition
\[
\Delta \Tij = \Delta \Tij ^{(1)} + \Delta \Tij ^{(2)}.
\]
Increments with superscript $(1)$ are associated, \emph{via} the contact law,
to the relative displacement of the contact point, which defines the constitutive part of the
stiffness matrix discussed in Section~\ref{sec:rist}, and
superscript $(2)$ labels increments of kinematic origin. 
We assume that the magnitude of the contact force is unchanged in the
absence of relative displacement at the contact ($\delta {\bf u}_{ij}=0$), and thus we write
\be
\Delta \Tij ^{(2)} = \Delta {\bf \theta}_{ij}\times \Tij ,
\label{eqn:deltatij2}
\ee
$\Delta {\bf \theta}_{ij}$ denoting the rotation of 
the contact region. $\Delta {\bf \theta}_{ij}$ can be split in a \emph{rolling} 
part $\Delta {\bf \theta}_{ij}^{(R)}$,
orthogonal to $\nij $, and a \emph{pivoting} one, $\Delta {\bf \theta}_{ij}^{(P)}$, along $\nij $.
$\Delta {\bf \theta}_{ij}^{(R)}$ is determined by the incremental change in $\nij $: 
\be
\Delta {\bf \theta}_{ij}^{(R)} = \nij \times  \Delta \nij . \label{eqn:dthetaR}
\ee
As to the pivoting part $\Delta {\bf \theta}_{ij}^{(P)}$, it is natural to equate it to the average
rotation of the two particles around the normal direction:
 \be
\Delta {\bf \theta}_{ij}^{(P)} = \frac{1}{2}
\nij \cdot\left(\Delta {\bf \theta}_i+\Delta {\bf \theta}_j\right) \nij . \label{eqn:dthetaP}
\ee
This choice is such that the rotation of the contact force coincides with the rotation of the pair in contact if both
objects move together like one rigid body (a condition of objectivity~\cite{KuCh06}).

Injecting~\eqref{eqn:deltan}
into Eqns.~\eqref{eqn:dthetaR} and~\eqref{eqn:dthetaP}, one readily obtains the appropriate formula for
$\Delta \Tij ^{(2)}$,
$$
\ba
\Delta \Tij ^{(2)}=&-\left[\Tij \cdot 
\left(\tilde {\bf u}_j -\tilde {\bf u}_i -\ww{\epsilon}\cdot {\bf r}_{ij}
\right) \right]
\frac{\nij}{r_{ij}}\\
&+\frac{1}{2} \left[\left(\Delta \theta _i+\Delta \theta _j\right)\cdot \nij\right](\nij \times \Tij).\ea
$$
With the notations of Section~\ref{sec:rist}, the contribution of contact $i,j$ to the $ 3\times 3$ diagonal block of 
$\stib$ which expresses the dependence of ${\bf F}_i^{\text{ext}}$ on displacement ${\bf u}_i$ is the non-symmetric
tensor
$$
-\frac{N_{ij}}{r_{ij}}(\ww{1}-\nij \otimes \nij)+\frac{\nij \otimes \Tij}{r_{ij}}.
$$
In general, the geometric stiffness matrix $\stib$ is thus not symmetric, except in frictionless sphere packings, which are
analogous to central force networks. We also note  that terms of order $N/R$ or $\normm{T}/R$ in $\stib$ correspond to terms of
order $K_N$ or $K_T$ in $\stia$, which are always very much larger. 

In the frictionless case, $\stib$ is a symmetric, negative matrix if forces are repulsive, 
as discussed by Alexander~\cite{SA98}. Any mechanism on
the backbone leads to an instability: the potential energy of the externally
applied load is strictly decreasing in that motion. This destabilizing effect can also be directly established in the rigid case,
as shown in~\cite{JNR2000}. This is the reason why stable packings of frictionless spheres in equilibrium under some
externally applied load are devoid of mechanisms involving the backbone.

In general stiffness matrices were discussed by 
Kuhn and Chang~\cite{KuCh06}, and by Bagi~\cite{Bagi07}. 
Those authors gave general results for $\stib$ with particles of arbitrary
shapes, which coincide with ours in the case of spherical balls.
\section{Stress-controlled molecular dynamics \label{sec:appendixcellm}}
It might be noted that the original Parrinello-Rahman equations slightly differ from ours. First,
Eqn.~\eqref{eqn:newtong} is written down with an additional term 
\be
m_i \ddot s_i^{(\alpha)} = \frac{1}{L_\alpha} F_i^{(\alpha)}-2\frac{\dot L_\alpha}{L_\alpha}m_i \dot s_i^{(\alpha)}.
\label{eqn:newtongtot}
\ee
Then, eqn.~\eqref{eqn:newtonc} is written with a different stress tensor, $\pk$, the definition of which
involves a particular reference value $\ww{L_0}$ of the cell dimensions, for which the volume is $\Omega_0$. $\pk$
is related to the Cauchy stress tensor $\ww{\Sigma}$ by
\be
\pk = \frac{\Omega}{\Omega_0} \ww{L}_0\cdot  \ww{L}^{-1} \cdot \ww{\Sigma}\cdot \trs{L}^{-1} \cdot
\trs{L}_0^{-1}.
\label{eqn:lagrangestress}
\ee
$\pk$ is a symmetric tensor known as the second Piola-Kirchhoff stress tensor~\cite{SAL01}
(also called \emph{thermodynamic tension} by some authors~\cite{RaRa84,LeMa06}).
Tensor $\pk$ can be used
to express the power $\Pp /\Omega_0$ of internal forces, per unit volume in the reference (undeformed)
configuration.

The \emph{metric tensor} $\GG=\trs{L}_0^{-1}\cdot \trs{L}\cdot \ww{L}\cdot \ww{L}_0^{-1}$
expresses the distance
between current points as a function of their coordinates in the reference configuration.
The difference between  $\GG$ and the unit 
tensor defines the Green-Lagrange strain tensor~\cite{SAL01}, $\ww{e}$, as
\be
\ww{e} = -\frac{1}{2}\left( \GG - \ww{I}\right).\label{eqn:defte}
\ee
Then $\pk$ is such that, for whatever strain history
\be
\Pp = \Omega_0 \pk : \ww{\dot{e}}.
\label{eqn:pp}
\ee 

If the last term of
Eqn.~\eqref{eqn:newtonc} is replaced by ${\displaystyle \frac{(L_0^\alpha)^2}{\Omega_0 L_\alpha}}$,
and if \eqref{eqn:newtongtot} is used instead of
\eqref{eqn:newtong}, then, in the case when interparticle forces derive from a potential $V$,
function of particle positions ${\bf r}_i$ and
orientations, the system of equations conserves the total energy 
\be
\ba
H = &\frac{1}{2} \sum_i m_i \dot {\bf s}_i \cdot \trs{L}\cdot  \ww{L}\cdot\dot {\bf s}_i 
+\frac{1}{2} \sum_i I_i \omega_i ^2 + V\\
+ &\frac{1}{2} M \cdot \trs{\dot L}:\ww{\dot L} + V + \frac{\Omega_0}{2} \GG :\pk .
\ea
\label{eqn:etot}\ee 
Such equations would tend to impose a constant Piola-Kirchhoff stress.

Granular assemblies are however dissipative systems, and energy conservation is not a crucial
issue as in molecular systems. In practice, we observed that omission of the extra term
of~\eqref{eqn:newtongtot} as well as control of Cauchy, rather than
Piola-Kirchhoff stresses, did not affect the approach to equilibrium configurations. Yet, on considering
small motions and elastic properties close to an equilibrium configuration, one 
may prefer dealing with external forces deriving from a potential. We note that this is indeed the case if we use
Eqns.~\eqref{eqn:newtongtot} and~\eqref{eqn:newtonc} with an isotropic $\ww{\Sigma}$,   
$\ww{\Sigma} = P\ww{1}$, in which case the external stress control is associated with potential energy $P\Omega$
(instead of the last term in Eqn.~\ref{eqn:etot}). 

\section{Comparison of local bond order parameters with experimental observations \label{sec:appaste}}
Although the samples were not isotropic, owing to the role of gravity in the preparation stage, and not perfectly homogeneous, because of
lateral walls, the X-ray tomography experiments by Aste \emph{et al.}~\cite{AsSaSe05} provided an unprecedented wealth of results on the geometry of
sphere packings. The local arrangement of neighbors around one bead 
were classified according to the values taken by the pair $(Q_4^{\text{loc}}(i),Q_6^{\text{loc}}(i)$, 
for different choices of the distance $d_c$ that defines the bonds. 
For each sample and choice of $d_c$, the proportion of spheres having the most frequent range of values
$(\hat Q_4 \pm 0.05,  \hat Q_6 \pm 0.05)$ corresponding to some typical disordered arrangement was recorded, as well as the frequency of occurrence of
values  $(0.191 \pm 0.05,  0.574 \pm 0.05)$ and $( 0.097 \pm 0.05,  0.485 \pm 0.05)$ respectively corresponding to fcc-like and hcp-like local ordering.
Those values are compared to the ones we observed in numerical samples of similar density in tables~\ref{tab:aste1} and~\ref{tab:aste2}
(we kept the same $(\hat Q_4, \hat Q_6)$ couple).
\begin{table}[!hbt]
\caption{\label{tab:aste1} Most frequently values $(\hat Q_4(i),\hat Q_6(i))$ observed in the measurements of \cite{AsSaSe05} in a dense sample
($\Phi = 0.640\pm 0.005$, called sample F in~\cite{AsSaSe05}), and fraction (\emph{dis.}) of
local configurations within interval $(\hat Q_4 \pm 0.05,  \hat Q_6 \pm 0.05)$ 
obtained in experiments and in numerical simulations with similar densities : A, A' and C.}
\begin{ruledtabular}
\begin{tabular}{cccccc}
Sample & $d_c$ & $(\hat Q_4, \hat Q_6)$&\emph{dis} (\%)& \emph{fcc} (\%)& \emph{hcp} (\%)\\ \hline
Aste \emph{et al.} & 1.1 &(0.23, 0.44)&31&6&4\\
dense& 1.2&(016, 0.45)&38&4&12\\
$\Phi \simeq 0.640$&1.3&(0.13, 0.42)&43&1&17\\
(F in \cite{AsSaSe05})&1.4&(0.10, 0.38)&47&3&13\\
\hline
A&1.1&(same&32&$<10^{-2}$&4\\
A&1.2&values&37&$<10^{-2}$&5\\
A&1.3&as&43&$<10^{-2}$&8\\
A&1.4&above)&46&$<10^{-2}$&12\\
\hline
A'&1.1&(same&31&$<10^{-2}$&5\\
A'&1.2&values&40&$<10^{-2}$&6\\
A'&1.3&as&45&$<10^{-2}$&10\\
A'&1.4&above)&48&$<10^{-2}$&15\\
\hline
C&1.1&(same&31&$<10^{-2}$&4\\
C&1.2&values&37&$<10^{-2}$&5\\
C&1.3&as&43&$<10^{-2}$&8\\
C&1.4&above)&46&$<10^{-2}$&13\\
\end{tabular}
\end{ruledtabular}
\end{table}
\begin{table}[!hbt]
\caption{\label{tab:aste2} Same as table~\ref{tab:aste1} for experimental samples
of lower densities, one with
$\Phi = 0.626 \pm 0.008$ (called sample D in~\cite{AsSaSe05}), another with $\Phi = 0.596 \pm 0.006$ (called sample B in~\cite{AsSaSe05}), to which values obtained
in simulated samples B and D, of similar densities, are respectively compared. }
\begin{ruledtabular}
\begin{tabular}{cccccc}
Sample & $d_c$ & $(\hat Q_4, \hat Q_6)$&\emph{dis} (\%)& \emph{fcc} (\%)& \emph{hcp} (\%)\\ \hline
Aste \emph{et al.} & 1.1 &(0.25, 0.44)&28&4&1\\
& 1.2&(0.19, 0.44)&35&2&7\\
$\Phi \simeq 0.626$&1.3&(0.15, 0.40)&42&1&11\\
(D in \cite{AsSaSe05})&1.4&(0.11, 0.36)&46&1&8\\
\hline
B&1.1&(same&30&$<10^{-2}$&2.7\\
B&1.2&values&38&$<10^{-2}$&7.6\\
B&1.3&as&43&$<10^{-2}$&10\\
B&1.4&above)&46&$<10^{-2}$&10\\
\hline
\hline
Aste \emph{et al.} & 1.1 &(0.30, 0.45)&24&3&1\\
loose& 1.2&(0.23, 0.44)&32&2&3\\
$\Phi \simeq 0.596$&1.3&(0.16, 0.38)&37&1&5\\
(B in \cite{AsSaSe05})&1.4&(0.14, 0.35)&43&2&5\\
\hline
D&1.1&(same&24&$<10^{-2}$&1.0\\
D&1.2&values&32&$<10^{-2}$&4.4\\
D&1.3&as&37&$<10^{-2}$&6.1\\
D&1.4&above)&43&$<10^{-2}$&7.5\\
\end{tabular}
\end{ruledtabular}
\end{table}
The fraction of beads with the typical disordered configuration of neighbors 
(marked \emph{dis.} in the tables) are very close in numerical packings and in the experimental one of the same densities, and
the frequency of occurrence of hcp-like local environments also compares well, although it does not seem to share the same dependence on the threshold distance
$d_c$ defining bonds. However, the small fraction of fcc-like beads observed in 
the laboratory is absent in the simulations. Many cirmcumstances can be invoked to explain
these differences, including of course the different packing history of the numerical and experimental samples, which in the the latter case 
involves gravity and anisotropy. It can be remarked once again that A' samples are a little more ordered than A ones
(with slightly larger fractions of hcp-like local configurations), from which
C samples are quite indistinguishable, as the quantities measured here do not depend on whether pairs of neighbors are actually in contact.
\section{Analysis of the free motion of divalent spheres \label{sec:appdival}}
We first give here the appropriate formulae to describe the free motion of divalent grains, then report on numerical stability tests.

The equations are specialized to the case of equal-sized spheres of dimater $a$, as in all the simulations reported in the present paper.
Let $i$ denote the label of the divalent grain in contact with its neighbors labelled $j$ and $k$. The  line joining the centers of
$j$ and $k$ is parallel to unit vector ${\bf e}$, defined as
$$
{\bf e} = \frac{\nik -\nij }{\normm{\nik - \nij }},
$$
and the distance $D$ of the center of $i$ to this line is 
$$
D = a\sqrt{1-({\bf e}\cdot \nij )^2}.
$$
$\omega _0$ denoting the angular velocity of the center of $i$ about this axis, the translational velocity of $i$, in its free
motion depicted on Fig.~\ref{fig:dival}, will be
\be
{\bf v}_i = \omega _0 D {\bf t},
\label{eqn:vi2}
\ee
the unit vector ${\bf t}$ being orthogonal to the plane containing the centers of $i$, $j$, and $k$:
$$
{\bf t}=\frac{\nij \times \nik }{\normm{\nik \times \nij  }}.
$$
Its angular velocity will be
\be
{\bf \Omega}_i = 2\omega _0 {\bf e}.
\label{eqn:omegai}
\ee
With such a choice, the instantaneous velocity of the (material) contact points between $i$ and $j$, or between $i$ and  $k$ satisfy:
$$
{\bf v}_i + {\bf \Omega}_i \times \frac{a}{2} \nij = {\bf v}_i + {\bf \Omega}_i \times \frac{a}{2} \nik = 0,
$$
as requested in a relative motion which is a combination of rolling and pivoting.

It is easy to check that the rules defined in Appendix~\ref{sec:appendixTrot} ensure that the tangential components of 
the contacts $i,j$ and $i,k$ rotate with the contact points around the axis joining the centers of $k$ and $j$ with angular velocity $\omega _0$.
In other words, the geometric stiffness matrix $\stib$ does not determine whether the mechanism associated with a two-coordinated bead is stable. 

We check for stability with numerical means, as follows. 
Starting from an equilibrium configuration, we first choose the potentially dangerous mechanisms, those involving relatively large contact forces, of the
order of the average normal force or even larger. Thus, grains $j$ and $k$ undergo significant changes in the moment of the contact
force with the mobile grain $i$. Then one such mobile divalent bead is attributed a velocity and a angular velocity according to 
Eqns.~\ref{eqn:vi2} and~\ref{eqn:omegai},
with a value of $\omega _0$ small enough for the centrifugal acceleration to be negligible (equal to $10^{-4} \normm{{\bf F}_{ij}}$). 
The evolution of the whole
packing under constant stress is then simulated with MD. Such numerical experiments were performed with the most fragile packings, D samples under low confining
pressures. In all cases studied, as expected, the motion of the mechanism 
entails very slow changes in the configuration, if any. 
The mobile grain maintains a constant angular velocity while nearly exactly following its circular trajectory for a long time, with hardly any change in kinetic energy.
These calculations are rather slow and costly, and we therefore limited our
investigations to 10 tests for 2 pressure levels in the D series, $P=1$~kPa and $P=10$~kPa. 
At the lowest pressure $P=1$~kPa, corresponding to $\kappa\simeq 181000$, these motions were often observed to
lead to a small rearrangement of the packing, in which kinetic energy spreads over all degrees of freedom, 
a significant fraction of the contacts, up to 25\%, go through a sliding stage, the contact network is slightly modified and the system restabilizes
in a slightly different configuration, with a small density increase (typically of order $10^{-5}$). In other cases, the freely moving grain stops when it collides with a
third grain other than the two with which it is maintaining contact. The system then finds a new equilibrium configuration without rearranging, only a few
contacts temporarily reach a sliding status. On repeating similar tests
at a larger confining pressure, $P=10$~kPa (still close to the rigid limit), 
the occurrence of this second scenario became much more frequent than the first, 
which was never observed in the 10 tests performed. 

\begin{figure}
\centering
\includegraphics*[width=8.8cm]{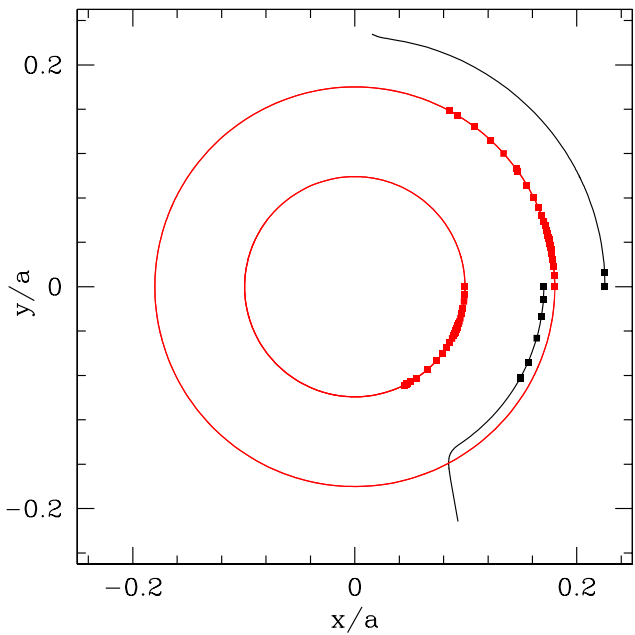}
\caption{\label{fig:divstab}(Color online) Examples of trajectories of mobile divalent grains in the plane orthogonal to vector ${\bf e}$.
 All trajectories begin at $x=D$, $y=0$.
Continuous lines depict 
the results obtained on ignoring contact creation, which end up, in 2 cases out of the 4 represented, in instabilities (black lines), while complete
circles are observed in the other, stable ones (red). Dots mark the same trajectories, which are in practice arrested by other contacts when contact creation is
taken into account.}
\end{figure}

The difference between stable and unstable cases is better appreciated 
on redoing the tests with a modified MD calculation method, in which only the initially existing
contacts are taken into account. Thus one only investigates the properties of the pre-existing contact network. 
If it breaks, the system globally falls apart, 
and nearly all contacts in the packing eventually open. This is the unstable case. In the stable case the mobile particle can turn
several times around the line of centers of its two contacting neighbors without notable changes in kinetic energy and the contact network is maintained.
This behavior is illustrated on Fig.~\ref{fig:divstab}, which displays trajectories of mobile grains in the plane orthogonal to ${\bf e}$. 
Note that such a procedure reveal instabilities that are prevented by
 the appearance of a third contact of the mobile grain, and hence overestimates the frequency of occurrence
of unstable configurations. 8 out of 10 such tests led to an instability in D samples under 1~kPa. 
This proportion fell to 2 out of 10 under 10~kPa.

Two possible conclusions may be drawn. On the one hand we may deem the D 
configurations under low pressures imperfectly stabilized, as some free motions might eventually cause
configurational changes. Or, since anyway the evolution is so slow, it may be pointed out, on the other hand, that the slightest amount of dissipation in rolling
would stop the free motion. In realistic systems the velocity a a free 
rolling motion always decays in time. 

Since the obtention of stable, equilibrated states in which all divalent grains have been made three-coordinated would be computationnally 
very costly, and as the occurrence of small instabilities related to such mechanisms appear to decrease fastly under growing confinement, we adopted the second 
attitude and regarded D and C configurations with a few percent of divalent particles as acceptable equilibrium states.

\end{document}